\documentclass[final,authoryear,5p,times,twocolumn]{elsarticle}

\usepackage{amsmath}
\usepackage{amssymb}
\usepackage{lineno,hyperref}
\modulolinenumbers[5]
\usepackage{fixltx2e}
\usepackage{afterpage}

\journal{Journal of \LaTeX\ Templates}

\biboptions{authoryear}

\usepackage{color}

\begin{document}

\begin{frontmatter}

  \title{An analysis method for data taken by Imaging Air Cherenkov Telescopes at very high energies under the presence of clouds}


\author[afil]{Sobczy\'nska Dorota}
\ead{dorota.sobczynska@uni.lodz.pl}
\author[afil]{Adamczyk Katarzyna}
\ead{kadamczyk@uni.lodz.pl}
\author[afil]{Sitarek Julian}
\ead{julian.sitarek@uni.lodz.pl}
\author[afil1]{Szanecki Michal}
\ead{mitsza@camk.edu.pl}
\address[afil]{University of {\L }\'{o}d\'{z}, 
 Department of Astrophysics, Pomorska 149/153, 90-236 {\L }\'{o}d\'{z}, Poland}
\address[afil1]{CAMK, ul. Bartycka 18, Warsaw 00-716, Poland}

\begin{abstract}
The effective observation time of Imaging Air Cherenkov Telescopes (IACTs) plays an important role in the detection of $\gamma$-ray sources, especially when the expected flux is low. This time is strongly limited by the atmospheric conditions. Significant extinction of Cherenkov light caused by the presence of clouds reduces the photon detection rate and also complicates or even makes impossible proper data analysis. However, for clouds with relatively high atmospheric transmission, high energy showers can still produce enough Cherenkov photons to allow their detection by IACTs. In this paper, we study the degradation of the detection capability of an array of small-sized telescopes for different cloud transmissions. We show the expected changes of the energy bias, energy and angular resolution and the effective collection area caused by absorption layers located at $2.5$ and $4.5\;{\rm km}$ above the observation level. We demonstrate simple correction methods for reconstructed energy and effective collection area. As a result, the source flux that is observed during the presence of clouds is determined with a systematic error of $\lesssim$20$\%$. Finally, we show that the proposed correction method can be used for clouds at altitudes higher than 5~km a.s.l.. As a result, the analysis of data taken under certain cloudy conditions will not require additional time-consuming Monte Carlo simulations.

\end{abstract}
\begin{keyword}
\texttt{$\gamma$-rays: general -- Methods: observational -- Instrumentation: detectors -- Telescopes}
\end{keyword}
\end{frontmatter}

\section{Introduction}
The successful use of the Imaging Air Cherenkov Technique in 1989 by the Whipple collaboration \citep{whipple} has allowed rapid development of ground-based $\gamma$-ray astronomy. The Cherenkov photons created in the atmosphere by relativistic charged particles, which are produced during the development of an Extensive Air Shower (EAS), are recorded by a matrix of photomultipliers located in the focal plane of the telescope. As a result, the two dimensional angular distribution of the Cherenkov light from an EAS (the shower image) is measured for each triggered event. The imaging method exploits the differences between images of hadron and $\gamma$-ray initiated showers in order to identify primary photons from the potential $\gamma$-ray source. The upcoming generation IACT array, called the Cherenkov Telescope Array (CTA) \citep{act11,ach13}, will achieve an exceptional sensitivity in the energy range between a few tens of GeV and a few hundred TeV. For this purpose, CTA plans to build telescopes in three sizes: large- (LST), medium- (MST) and small-sized telescopes (SST).
  
The atmosphere is an integral part of IACTs. First, the amount of Cherenkov light produced depends on the atmospheric profile, i.e., dependence of the refraction index on the altitude. Second, Cherenkov light is scattered and absorbed by the atmosphere before reaching the observation level. The second effect is much stronger if clouds are present during observations.
Therefore, atmospheric conditions are monitored during data taking in experiments such as H.E.S.S. \citep{aha06,devin}, MAGIC \citep{ale12} and VERITAS \citep{weekes2002,hold11}. Additional instruments to measure the transparency of the atmosphere are also planned to be built for CTA \citep{doro13,lop13,gaug16,iarl16,valore17}. As an example, the LIDAR system used in MAGIC can resolve narrow cloud layers of even $100-200\; {\rm m}$ (cf. Fig.3 in  \citealt{fruck13}) in time scale of the order of minutes. Similar or better performance is expected from Raman LIDAR systems used in CTA (cf. \citealt{gaug18}, Fig.4). 

The lateral density distribution of the Cherenkov light strongly depends on the atmospheric profile \citep{bern00, bern13}. The presence of clouds influences the data in two ways.
First, when the reduced number of Cherenkov photons hits the reflector, the amount of light can become too small to trigger the telescope or to reconstruct the shower. As a result, both the detection rate and the effective collection area decrease, leading to an increase of the energy threshold \citep{rult13}.
Second, the shower images may be deformed. This deformation worsens shower reconstruction and degrades the $\gamma$/hadron separation efficiency \citep{sob14}.
For fixed zenith angle, the effect of clouds on the data depends not only on the transparency and altitude of the clouds but also on the type of primary particle, its energy and the impact parameter \citep{sob15}.

A method to correct the reconstructed energy of the shower and fluxes of primary particles due to the presence of low-level aerosols has been already studied in \citep{nolan10,devin} for the H.E.S.S. data.
The authors show that the effective collection area (as a function of reconstructed energy) is reduced for low-level clouds in comparison to clear sky simulations. Therefore, by including the atmosphere with low-level aerosols (detected by LIDAR) in the Monte Carlo simulations, the effect of additional atmospheric extinction can be corrected for primary energies below $10\; {\rm TeV}$ \citep{devin}. Additionally, both \citet{nolan10} and \citet{devin} showed that a bias in the reconstructed energy is expected if the energy is reconstructed based on the Monte Carlo simulations for cloudless conditions. It has been also shown in \citet{hahn14}, that in case of large IACTs and low-level extinction layers, the parameter describing the transparency coefficient can be estimated from the real background data itself without taking into account the LIDAR measurement. In this case, the underestimation of the flux normalization strongly depends on the transparency coefficient, e.g. for the transparency of 0.6, the flux normalization is only half that of the clear sky (see Fig.4 in  \citealt{hahn14}). A similar method of the monitoring the atmospheric transmission based on the background FACT data has been shown in \citep{hildebrand13,hildebrand17}.   

Data taken by MAGIC during nights with additional atmospheric extinction are analyzed in a special way, which includes corrections for the reduced atmospheric transmission \citep{dorn09,garrido13,fruck13,fruck15}. For the dust layer (the Saharan Air Layer, known as the calima, or low-level clouds) that is below $5.5\; {\rm km}$, the shower maximum is well above this layer for energies below $1\; {\rm TeV}$.
Therefore, in the case of MAGIC-1 observations taken during calima with the extinction lower than 40$\%$, only the absolute light calibration has been corrected, as the image deformation is relatively low \citep{dorn09}.
Finally, an accurately reconstructed primary energy has been obtained based on the recalculated image parameters. The aerosol transmission, which is obtained from LIDAR data, is used in a simple but efficient correction method for MAGIC \citep{fruck13,fruck15}. It has been shown in those papers, that based on two assumptions, the source flux can be reproduced from the data taken in the presence of clouds. First, the reconstructed energy  is scaled up using the aerosol transmission folded with a normalized, energy and zenith angle-dependent, light emission model around the reconstructed shower maximum. The latter has been obtained from Monte Carlo simulations. 
Second, the expected collection area for the corrected energy is assumed to be the one evaluated at the energy before scaling up.
For relatively low energies ($<\sim$10~TeV) and low-level extinction layers ($<$7~km) the shower maximum lies above the cloud. Thus, images are mainly diminished, as most of the shower has already developed before reaching the cloud.

However, for higher primary energies, shower maxima may lie at or below the cloud altitudes. This may result in a significant distortion of the shower image (as the shower is not uniformly affected). The image is less deformed for small impact parameters where most of the detected Cherenkov light comes from heights close to the observation level. It has been shown \citep{sob15}, that the strength of this effect depends on the transparency and altitude of the clouds. The deformation is stronger for higher primary energies. 
The $\gamma$/hadron separation capabilities at high energies for the observation with clouds has already been studied \citep{sob14}. It has been presented in that paper that using scaling factors for the image parameters derived from clear sky simulations leads to a stronger degradation of the quality of the $\gamma$-ray separation than adapting those factors from the simulation with the presence of clouds.

Fluxes of $\gamma$-ray sources at very high energies (more than a few TeV) are low. Therefore, for the cosmic $\gamma$-ray detection a long observation time is required or a large number of IACTs has to be used at the same time. In CTA, the second strategy will be exploited, but also the effective observation time should be as long as possible. The main aim of our paper is to demonstrate that the duty-cycle of the SSTs can be increased by taking observations under cloudy conditions, while maintaining similar measurement accuracy. As the example, we have simulated a small array of small-sized telescopes with one mirror (SST-1M) \citep{barnacka13,heller17,sliusar17}

In this work, we used simulations of observations with and without cloud cover. 
We developed a simple analysis method that can be used without generating dedicated Monte Carlo simulations for different cloud heights and transmission.
We evaluate the performance of SST-1M array observations with clouds and investigate the systematic uncertainties of our method on flux reconstruction. 

\section{Description of the Monte Carlo simulations}
\begin{figure}[t]
\begin{center}
\includegraphics*[width=7cm]{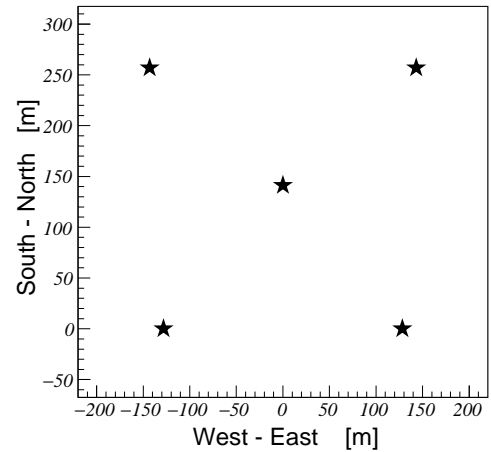}
\end{center}
\caption{The layout of the telescope system on the ground. The side of the square is approx. $260\; {\rm m}$}
\label{fig:geom}
\end{figure}

\begin{center}
\begin{table}[t]
\caption{ Overview of parameters and the number of simulated events.}
\label{tab:tab1}
\begin{tabular}{|l|l|l|}
\hline
primary particle &$\gamma$&proton\\
\hline
E$_{min}$[GeV]&300&800\\
\hline
E$_{max}$[TeV]&150&450\\
\hline
impact$_{max}$ [m]&1100&1600\\
\hline
View Cone [$^{\circ}$]&0&12\\
\hline
number of events&10$^{8}$&2.3$\cdot$10$^{8}$\\
\hline
\end{tabular}
\end{table}
\end{center}

We have simulated the development of an extensive air shower using the CORSIKA code \citep{heck, knapp}, version 6.990. The UrQMD {\citep{urq1,urq2}} and QGSJET-II-03 \citep{ostap06a,ostap06b,ostap07} interaction models have been applied for the low (a momentum of particle $<$ 80 GeV/c) and higher energy ranges, respectively. The simulations have been performed for the Armazones site in Chile, located at 2.5~km above sea level. This site is close to Paranal site in the Atacama Desert in Chile (i.e. the final selected site for CTA-South). The altitude of Paranal is 400~m less than Armazones, which may result in a slight change in the performance of IACTs array for the investigated energy range. It should also be noted that the atmospheric conditions for Paranal site are very good because 96$\%$ of the nights are clear \citep{Kurlandczyk2007}, but the presented method of analysis can also be used for other IACTs locations.

As example of an SST array we have chosen a set of five IACTs. The layout of the telescope systems simulated is presented in Fig.~\ref{fig:geom}. The distance between the two closest IACTs in our telescope configuration (approx. 185~m) is comparable with inter-telescope distances for the best performing CTA-South layout (they varied between 190 and 300~m) \citep{ach19}. Due to the fact that the simulation of the shower development is time consuming for the investigated energy range, we reused the same shower 20 times - the full IACTs array has been shifted with respect to the shower's axis core position. A fixed direction of the simulated primary $\gamma$-rays has been chosen to be $20^{\circ}$ in zenith and $0^{\circ}$ in azimuth angles (showers pointing to the North). The proton-induced showers were simulated within a cone with a half-opening angle of $12^{\circ}$ at the same zenith and azimuth. The overview of parameters and the numbers of simulated events (after re-usage) are presented in Table ~\ref{tab:tab1}. The value of impact$_{max}$ corresponds to the maximum impact parameter from the point with coordinates (0,0) on Fig.~\ref{fig:geom}.

The sim$\_$telarray code \citep{bern08} (with the settings of the CTA prod3) has been used for the telescope simulations. The atmospheric extinction has been taken into account in the detector simulation. The atmospheric extinction coefficients for Armazones site (which are part of the sim$\_$telarray package) were used for the simulation of cloudless conditions. The additional extinction due to the presence of clouds has been taken into account based on the formulas of the extinction coefficients of a cloudy medium presented in \citep{kokh08}. Two altitudes, 5 and 7~km a.s.l. (the height of the bottom layer), of clouds with thickness of $500\; {\rm m}$ were studied in this paper. Those heights correspond to the average positions of the vertical shower maximum for $\sim$100~TeV and $\sim$10~TeV, respectively. We have not simulated very high clouds because for energies above 5~TeV even fully opaque clouds at 10~km have small influence (the order of $\sim$10$\%$) on the Cherenkov light density (see Fig.1b and Fig.2a in \citealt{sob13, sob14}).

The different water concentrations in cloudy media were chosen in order to get the total cloud transmission (T) of 0.8, 0.6, 0.4 and 0.2 \citep{adam16}. The total transmission is wavelength dependent in the formula from \citep{kokh08}.  We have obtained that the transmission at 1000~nm is lower by only 2-3$\%$ than at 200 nm. The normalization of T to the mentioned-above values has been done for $200\; {\rm nm}$ photons. Additionally, we have simulated a cloud at 6~km a.s.l. with the total transmission of 0.7 and we used this MC simulation set to test the analysis method of the data taken in the presence of clouds. 

For the image cleaning\footnote{We applied the so-called 2-pass image extraction at levels $8-4\; {\rm phe}$ and $5-2.5\; {\rm phe}$.} and parameterization, and the estimation of stereo parameters and the primary energy, we used the MARS/Chimp chain \citep{zan13,ale16,sit18}. In particular, the energy estimation is done with the help of the Random Forest \citep{alber2008} for each telescope separately and then averaged with weights. The same tool has been used to select primary $\gamma$-rays from the protonic background. 

\begin{figure}[!htb]
  \centering
\includegraphics[width=0.45\textwidth]{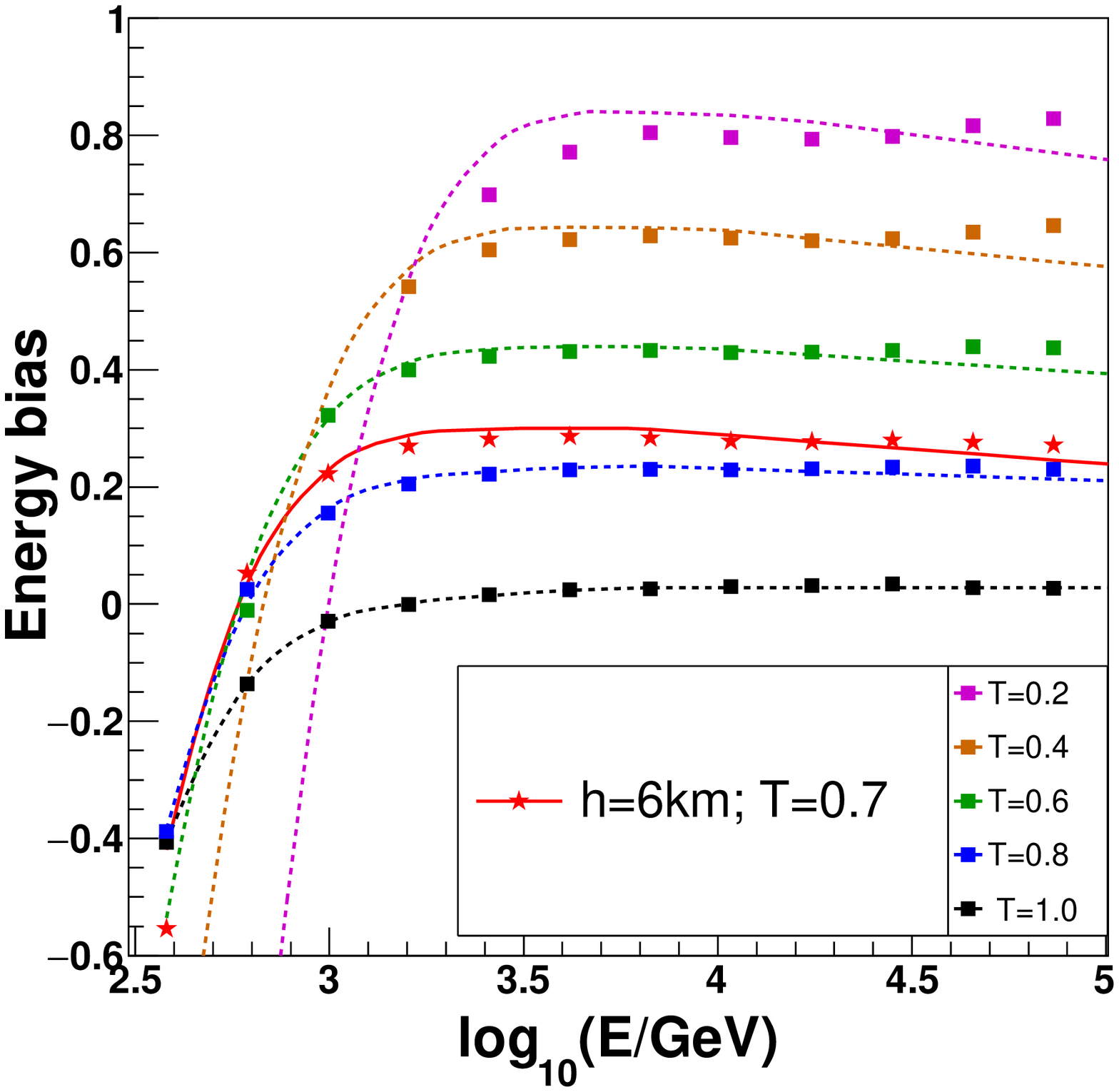}
\includegraphics[width=0.45\textwidth]{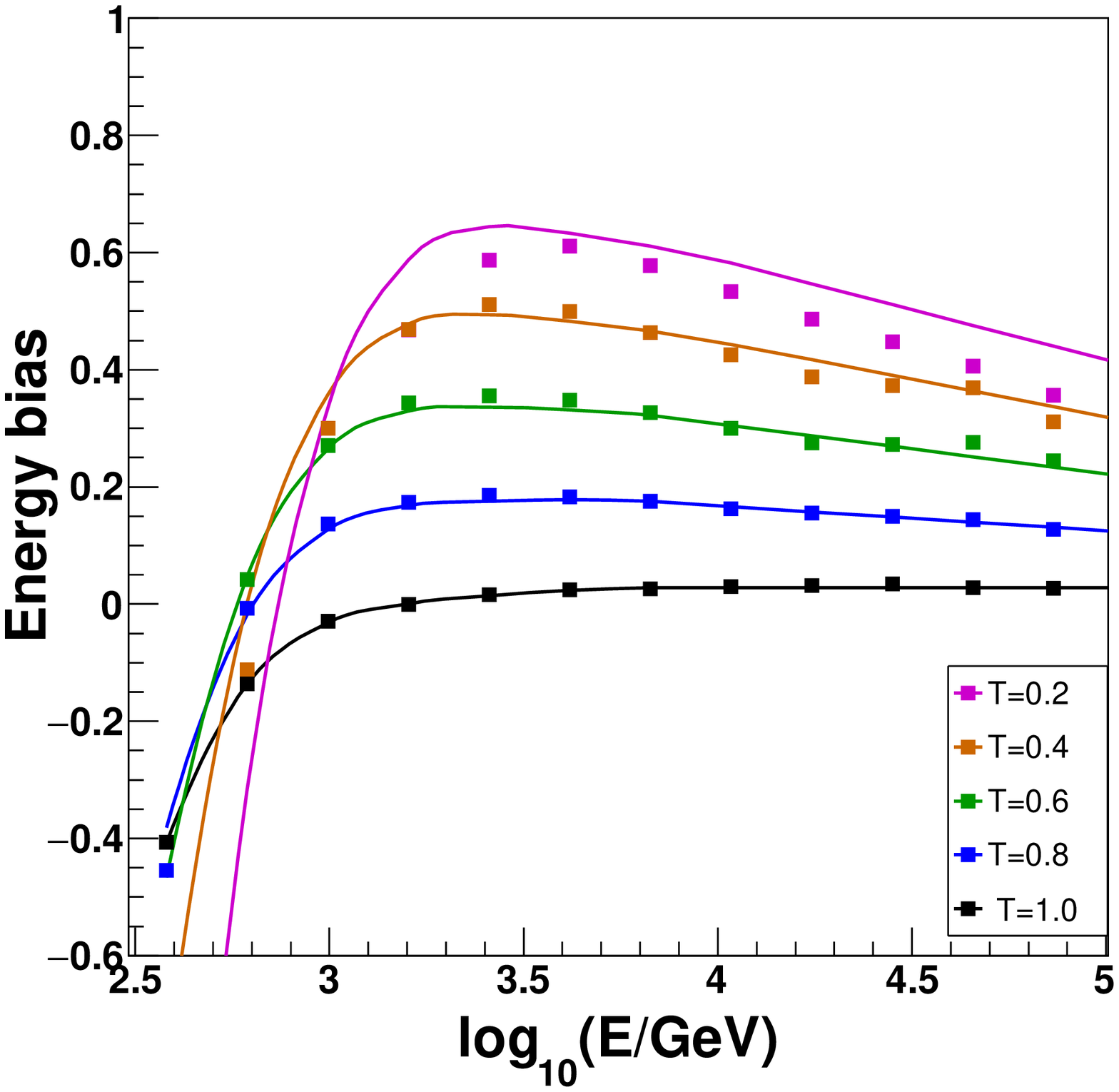}
\caption{The energy bias $b$, see Formula \ref{eq:bias1}, of reconstructed $\gamma$-rays versus the primary energy for a cloud at: 5~km and 6~km - top panel; 7~km -bottom panel. Black lines present the results of clear sky simulation. Different colors correspond to the different cloud transmission (see description in the legend). Points show the results of the simulation, while lines are an approximation given by equation~\ref{eq:bias2}.}
\label{fig:bias}
\end{figure}

\section{Results and discussion}

In order to avoid an over-training in the reconstruction and $\gamma$/hadron separation procedure, we divide the MC simulations into a few subsamples. At first, one subsample of cloudless $\gamma$-ray data has been used for training the reconstruction of the energy and stereo parameters. Next, we applied it to subsamples of $\gamma$-ray and proton results for different cloud transmission. We defined an energy bias $b$ as a relative difference between a true and reconstructed energy, see Formula \ref{eq:bias1} below. In each true energy bin the mean of the Gaussian fit of the central part of the energy bias distribution has been found. The energy biases obtained in this way are presented as points in Fig.~\ref{fig:bias}.
For energies below 1~TeV the energy bias is negative for cloudless data (see black lines in Fig.~\ref{fig:bias}). The negative bias resulting from an overestimation of the reconstructed energy (below and close to the energy threshold) can be explained by two effects: the threshold effects and a lack of the simulations below 300~GeV.

In the case of a cloud at 5~km, i.e. below the shower maximum, the expected energy biases are approximately equal to (1-T) for all investigated T (see top panel of Fig.~\ref{fig:bias}). Most of the Cherenkov light is produced above the cloud and finally the image \textit{Size} parameters (defined as a sum of the signals from pixels that survive image cleaning) are smaller in comparison to the clear sky data (in the ideal case by a factor of (1-T)). Moreover, for more opaque clouds the low energy showers cannot trigger the IACT array or their energy cannot be reconstructed. The lack of points in Fig.~\ref{fig:bias} is caused by such effects where, due to the lack of statistics, the bias cannot be determined. When a cloud is above the shower maximum (i.e. at 7 km a.s.l., as shown in the bottom panel of Fig.~\ref{fig:bias}), only part of the shower image is absorbed by the cloud. This results in the situation that only a part of the shower image is absorbed by the cloud. The reduction of the \textit{Size} parameter depends on both the impact parameter and the primary energy (as the altitude of the shower maximum is energy dependent). For T$<$1 the bias decreases with energy above the energy threshold, which depends on T. For lower energies, the threshold effects are dominant and the bias increases with energy.

We propose an approximation of the energy bias based on the fractions of photons created above the cloud to all produced photons that hit the ground at distances larger than 80~m from the shower axis (see Appendix A). The fraction ($F_{ab}(E,H)$) is a function of energy and the cloud altitude (H). The natural assumption is that Cherenkov photons above the cloud participate in the energy reconstruction with a weight equal to the cloud transmissivity, while the light created below has a weight equal to 1. We define:

\begin{equation}
  \begin{aligned}
    b(E,T,H) \equiv \frac{E- E_{rec}(E,T,H)}{E}=(1-T)\cdot F_{ab}(E,H)
  \end{aligned}
  \label{eq:bias1}
\end{equation}
where $b(E,T,H)$ is an energy bias, $E$ and $E_{rec}(E,T,H)$ are the true and reconstructed energies.
\footnote{Note that for simplicity we exploited that the simulations were performed at low zenith angles, i.e. the transmission seen by Cherenkov photons is nearly equal to the vertical transmission of the cloud. In case the method is applied to observations at a high zenith angle $Zd$, $T$ should be substituted by $T^{1/cos(Zd)}$.}

Formula~\ref{eq:bias1} does not take into account the effects close to the threshold (like negative bias for T=1) and that is why we need to add the energy bias for cloudless conditions $b(E\cdot\tau (E,T,H),1,0)$, where $\tau (E,T,H)$ is the total atmospheric transmission for the cloud at altitude of H with transparency of T.

\begin{equation}
  \begin{aligned}
    \tau(E,T,H) \equiv T \cdot F_{ab}(E,H)+1 \cdot (1-F_{ab}(E,H))\\
    = 1 - (1-T)\cdot F_{ab}(E,H)
  \end{aligned}
\label{eq:tau}
\end{equation}

Furthermore, to improve the agreement between the cloud simulated in MC and our formula we had to add a correction factor (A). Constant $A$ is a factor that depends of the chosen IACT array. In our case A equal to 1.2 was applied to all simulated clouds and it fits to MC results for the cloud transmission higher than 0.4. Finally the energy bias can be described by:

\begin{equation}
  \begin{aligned}
    b(E,T,H) = A \cdot (1-T) \cdot F_{ab}(E,H)+b(E \cdot \tau (E,T,H), 1, 0)\\
    = 1 -  \tau_{A} (E,T,H)+b(E \cdot \tau (E,T,H), 1, 0)
  \end{aligned}
\label{eq:bias2}
\end{equation}
where $\tau_{A}(E,T,H)$ is the corrected total atmospheric transmission for gamma showers with energy E (see Appendix A), which includes a constant A.

All curves in Fig.\ref{fig:bias} are the results of our approximation~\ref{eq:bias2} and they agree with the data within $\lesssim10\%$ for transmission equal or higher than 0.6 and 0.4 for cloud altitudes of 5 and 7~km a.s.l, respectively. Additionally, we show that the formula for the energy bias can be used for other clouds without changing the factor A. In Figure \ref{fig:bias} we plot the bias predicted by Equation  \ref{eq:bias2} for a cloud at 6 km a.s.l. with T=0.7 (red curve), as well as the points obtained from MC simulations of these conditions (red stars). 

\begin{figure}[t]
\begin{center}
\includegraphics*[width=7cm]{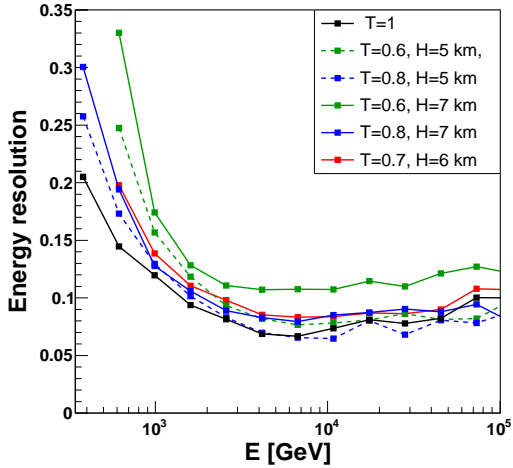}
\end{center}
\caption{The energy resolutions versus the primary energy for clear sky (black solid line) and the presence of clouds (see legend for).}
\label{fig:eresol}
\end{figure}

\subsection{Energy resolution}

For large energy biases, the standard definition of the energy resolution is not a useful metric for assessing performance. In the following analysis, a corrected energy is used instead of the reconstructed one. Therefore, based on the approximation \ref{eq:bias2}, the corrected energies ($E_{cor}$) were obtained and an energy resolution can be defined as a standard deviation of an $[(E-E_{cor})/E]$ distribution for a given true energy bin. Figure~\ref{fig:eresol} shows the energy resolutions obtained for cloudy and clear sky conditions (black solid line). Note that for the cloud transmissions $\geq$ 0.6 and E$>\sim$2~TeV the energy resolution is lower than 13$\%$ at its plateau, while in a case of the cloudless conditions the energy resolution is smaller than 9$\%$. For energies below 2~TeV the threshold effects worsen the energy resolution. It is worth noticing that, even for cloudless conditions, for low energies the energy resolution strongly depends on the energy, as the relative fluctuations of the Cherenkov light density are energy dependent \citep{chitnis98,sob09}. For E$>$2~TeV i.e. much above the threshold, statistical uncertainties of the corrected energy are small ($<$15$\%$) for a cloud transmission above 0.6.

\subsection{Angular resolution}

\begin{figure}[!htb]
  \begin{center}
    \includegraphics[width=0.4\textwidth]{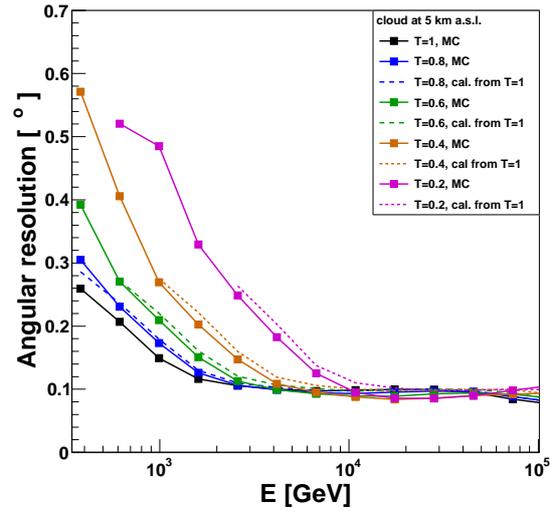}    
    \includegraphics[width=0.4\textwidth]{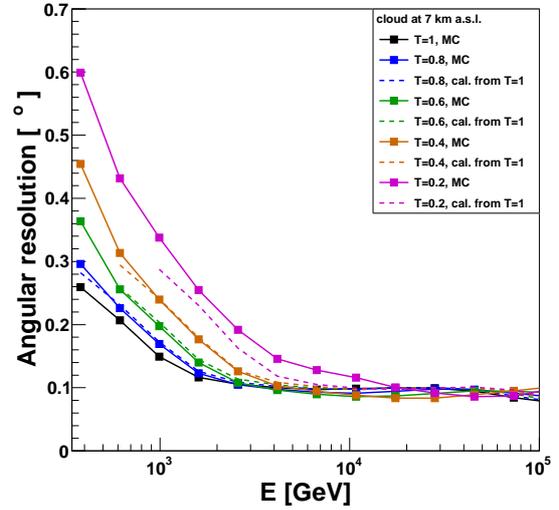}
    \caption{The angular resolution versus the primary energy for clouds at: 5~km - top panel; 7~km a.s.l. - bottom panel. All points and solid lines correspond to the results of the MC simulations with the presence of clouds, while dashed lines present the angular resolution for different cloud transparencies that have been obtained for the results of the clear sky by using formula~\ref{eq:angular}}     
    \label{fig:angular}
    \end{center}
\end{figure}

The angular resolution is defined as the radius of a circle containing 68$\%$ of all reconstructed events, which have the angular distance between the simulated and reconstructed directions smaller than this radius, for primary $\gamma$ rays. Figure~\ref{fig:angular} shows how the angular resolution changes for observations in the presence of clouds. The angular resolution for clear sky improves with the energy and finally almost stabilizes at the level of 0.1$^{\circ}$. The presence of clouds with T$\geq$0.4 causes an increase of the angular resolution only for energies lower than $\sim$4~TeV. The worsening of the direction reconstruction becomes severe as the cloud transmission diminishes. For a cloud transmission equal to 0.2 the angular resolution stabilizes for energies higher than 10~TeV.

In order to prove that data taken under cloudy conditions can be analyzed using only the simulation of cloudless sky, we checked that the expected angular resolution curve for T$<$1 can be reproduced using simulations of a fully transparent atmosphere. The most natural and simplest approximation is a scaling of the energy by using the corrected total atmospheric transmission (see Appendix A):

\begin{equation}
  \sigma_{\theta}(E,T,H) \equiv \sigma_{\theta}(E\cdot \tau_{A}(E,T,H),1,0)
  \label{eq:angular}
\end{equation}
where $\sigma_{\theta}(E,T,H)$ is the angular resolution for the energy of E in case of the cloud at altitude H and the transmission of T. All dashed lines in Figure~\ref{fig:angular} show the results of the equation~\ref{eq:angular}. The simple energy scaling makes it possible to estimate the angular resolution accurately enough for cloud transmissions higher than 0.2.

\begin{figure}[!htb]
\begin{center}
\includegraphics[width=0.45\textwidth]{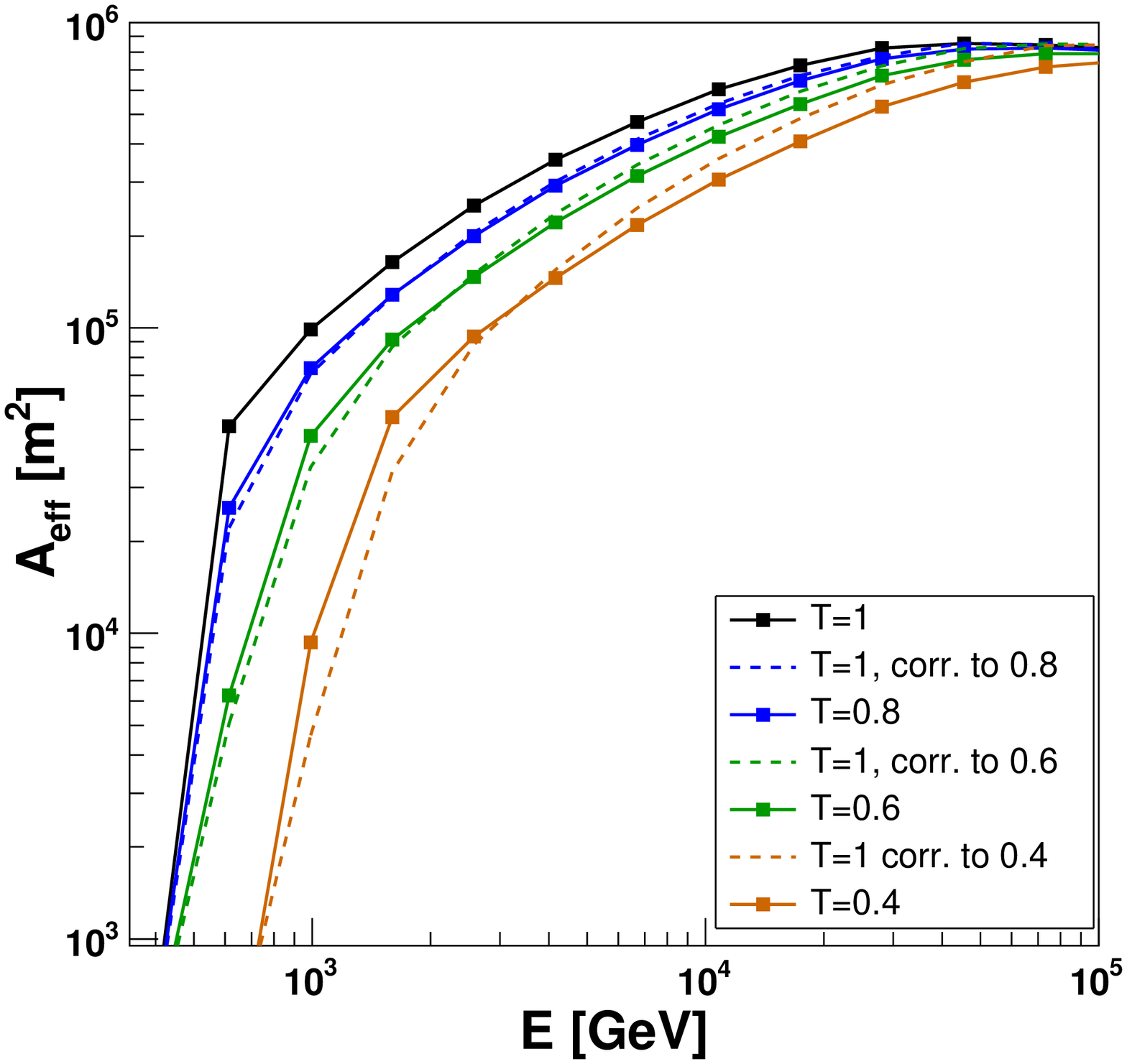}
\includegraphics[width=0.45\textwidth]{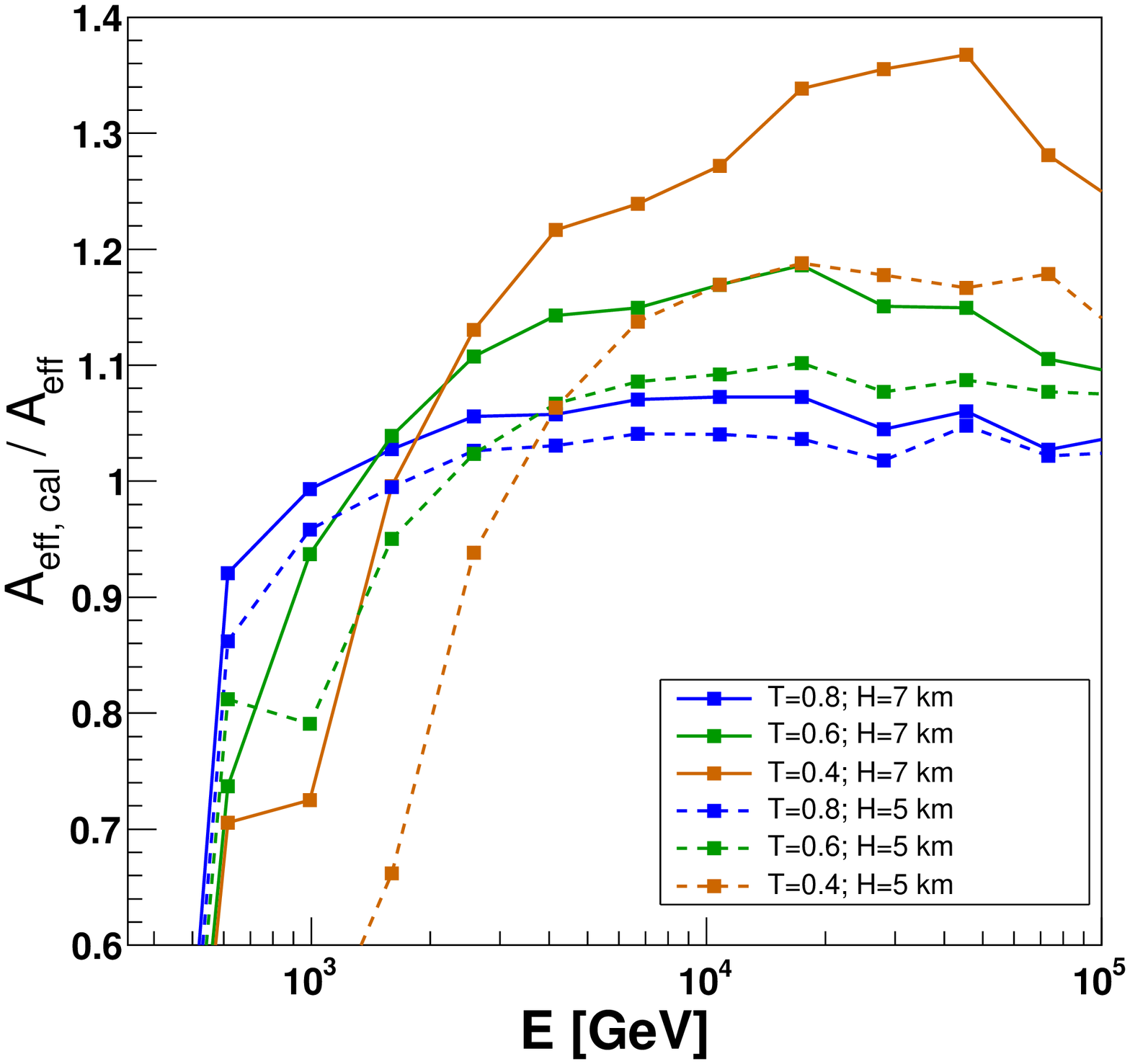}
\caption{Top panel: The effective collection area after reconstruction as a function of the energy for clouds at 5~km. All points and solid lines correspond to the results of the MC simulations, while dashed lines present the expected collection areas for different cloud transparencies that have been obtained using formula ~\ref{eq:effcollarea}. Bottom panel: The ratios between the calculated $S_{eff, cal}(E,T,H)$ and effective collection area obtained from the full simulation of the cloud.}
\label{fig:collarea}
\end{center}
\end{figure}

\subsection{Effective collection area}

The effective collection areas ($A_{eff}$) after reconstruction versus the true energy are shown as solid lines on the top panel of Figure~\ref{fig:collarea} for a cloud altitude of 5~km a.s.l. The degradation of the collection area in the presence of clouds is mainly caused by the decrease of the trigger rate due to the lower Cherenkov photon densities. As high energy events observed with clouds imitate lower energy showers, we propose to estimate the collection area for data taken in the presence of clouds based on results of the clear sky simulations by simple energy scaling:

\begin{equation}
  A_{eff,cal}(E,T,H) \equiv A_{eff}(E\cdot \tau_{A}(E,T,H),1,0)
  \label{eq:effcollarea}
\end{equation}

The effective collection areas in the presence of clouds that were calculated using formula~\ref{eq:effcollarea} are presented on the top panel of Figure~\ref{fig:collarea} as dashed lines.

The ratios between the calculated $A_{eff, cal}(E,T,H)$ and effective collection area obtained from the full MC simulation of the cloud are shown on the bottom panel of Figure~\ref{fig:collarea} as solid and dashed lines for cloud altitude of 7 and 5~km, respectively. For energies above a few TeV, formula ~\ref{eq:effcollarea} overestimates $A_{eff}(E,T,H)$, which would result in an underestimation of the flux from the source. However, for cloud transmission equal or above 0.6 the relative systematic error caused by using formula~\ref{eq:effcollarea} is lower than 20$\%$. Note that for energies above a few ~TeV the scaling formula works better for lower than for higher cloud.

For low energies the estimated collection area is lower than expected from MC. Thus, the positive bias in the reconstructed flux may be expected. It should be noted that in order to determine the spectrum, the effective collection areas after $\gamma$/hadron separation are used.
	
 \subsection{Gamma/hadron separation}
\begin{figure}[!htb]
  \begin{center}
\includegraphics[width=0.45\textwidth]{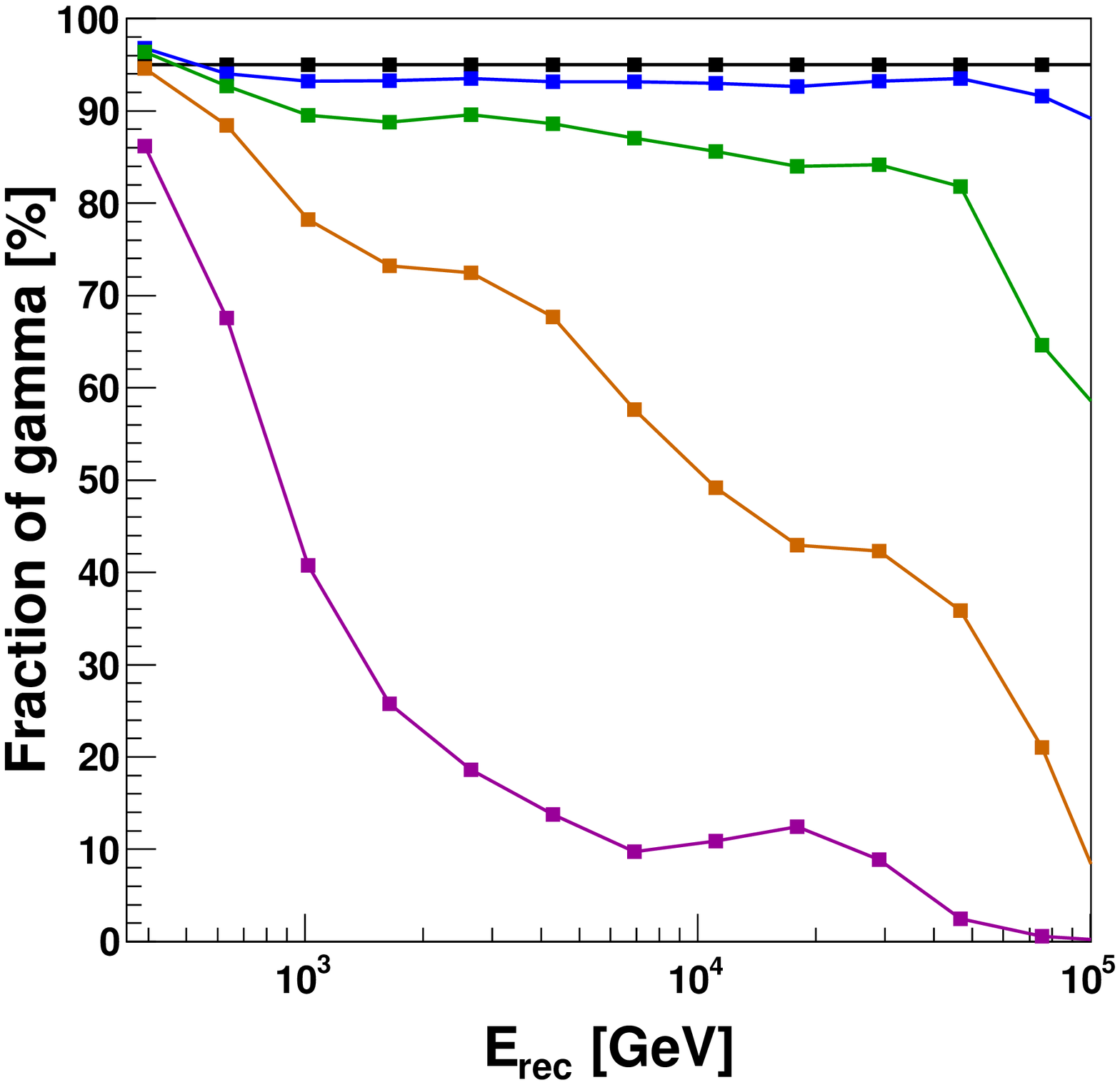}
\includegraphics[width=0.45\textwidth]{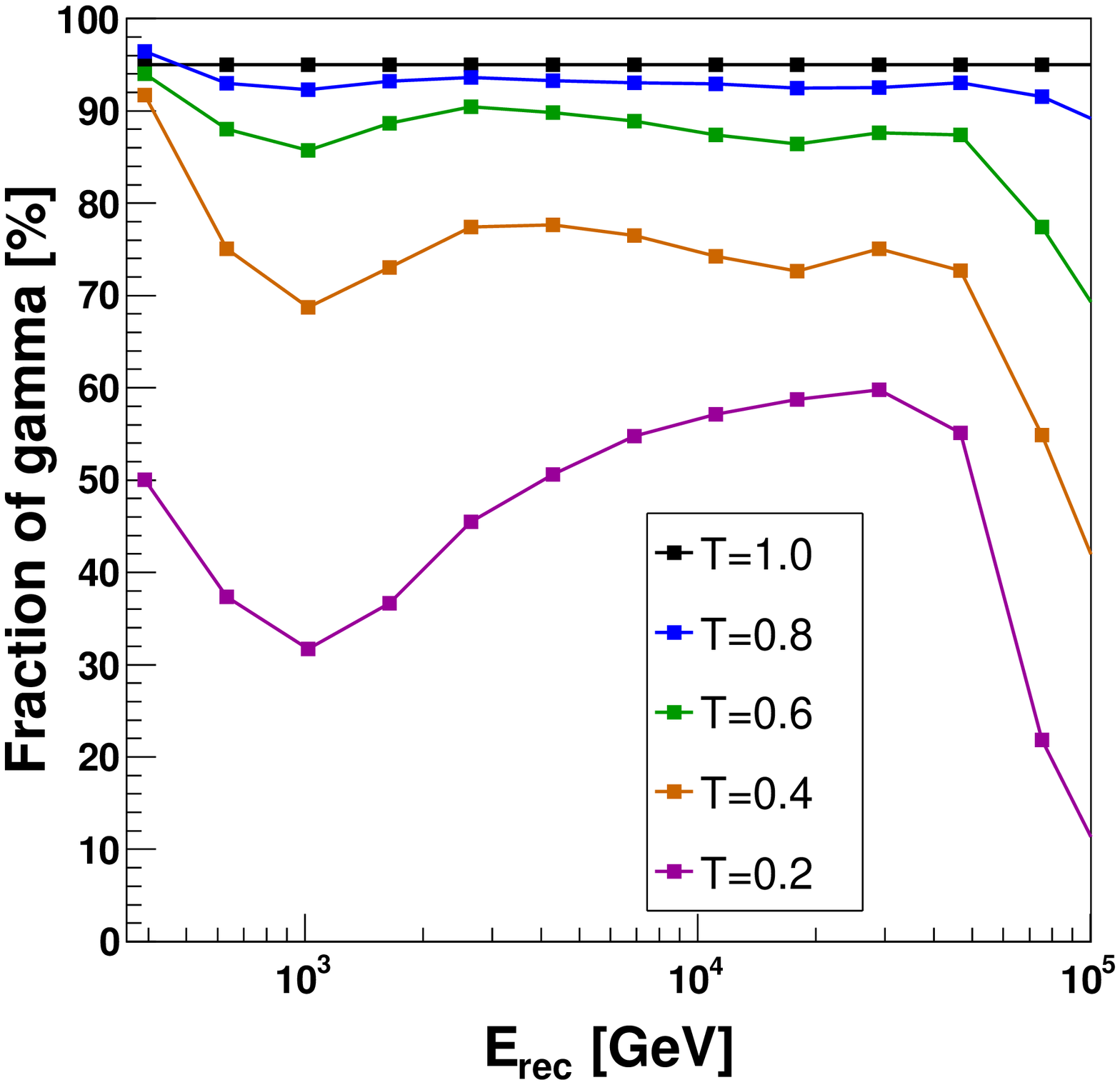}
\caption{The fraction of reconstructed $\gamma$-rays that survive G95 $Hadroness$ cut versus the reconstructed energy for a cloud at: 5~km - top panel; 7~km - bottom panel. Black line presents the results of clear sky simulation. Different colors correspond to the different cloud transmission (see description in the legend).}
\label{fig:fractionG95ii}
  \end{center}
\end{figure}

 The Random Forest method \citep{alber2008}, implemented in Mars/Chimp \citep{zan13,ale16,sit18}, was used for the selection of the primary gamma rays from the protonic background. We trained the Random Forest using subsamples of Monte Carlo simulations of proton- and gamma-initiated showers for clear sky conditions and next applied it to the subsamples of MC data which also include the additional extinction of the cloud. Finally, a $Hadroness$ parameter, which is determined on the basis of image parameters as well as stereo parameters, was assigned to each reconstructed event. The $Hadroness$ orders the events from the most to the least $\gamma$-like, hence a selection on hadronness increases the $\gamma$-ray purity of a sample.
 For later analysis we use a loose cut in $Hadroness$ (G95) that requires 95$\%$ of the reconstructed gamma-ray events to survive this cut at each energy bin for cloudless atmospheric conditions. The selected cuts for clear sky were applied to the data with additional extinction. Figure~\ref{fig:fractionG95ii} shows how the fraction of surviving gamma events depends on the reconstructed energy for different cloud transmissions. For a cloud altitude of 5~km the fraction of gamma events after G95 is smaller than for 7~km, as expected. Note that clouds transparencies lower than 0.4 and 0.6 (at 5 and 7~km, respectively) cause a significant degradation of the cut efficiency, i.e., less than 80$\%$ of $\gamma$-rays are selected as gamma-like events using very loose $Hadroness$ cuts (see the high energy part). Additionally, for low energies we have obtained significantly deteriorated angular resolution, thus a significant fraction of gamma rays will be excluded from the on-source sample by the $\theta^{2}$ cut.  The $\theta^{2}$ parameter is the square of the angular distance between the reconstructed and true directions of the shower.
 Taking into account the effectiveness of the $Hadroness$ cut, we decided to limit our analysis to clouds with a transparency of $\geq$ 0.6, because in such conditions over 80$\%$ of the true $\gamma$-ray events meet the gamma selection criteria, except for showers with E$>$50~TeV. Moreover, also the reproduction of the angular resolution, energy resolution and the effective collection area (based on the results of the clear sky  simulations) is working well in this limit. In the case of $A_{eff}$, the possible systematic overestimation of $A_{eff}$ is not higher than 20$\%$.      
 
Finally, based on the $\theta^{2}$ distributions we have chosen a fixed value for the $\theta^{2}$ cut of 0.025~deg$^2$ for all energies.
 
This cut is applied to the MC events for all analyzed total transmissions and altitudes of the cloud.

\subsection{Correction for atmospheric extinction}
An analysis method for the data taken in the presence of clouds has been shown in \citep{fruck13,fruck15}. In order to reconstruct the source spectrum, for each event recorded and reconstructed as gamma-like, the energy correction is applied based on the average optical transmission (the aerosol transmission profile obtained from the measurement of the real atmospheric condition, folded with an air-shower light emission profile assumed for the measured shower maximum of the same event). Next, all events in a single bin of corrected energy are summed up with a weight that is the inverse of the effective collection area obtained from the simulations for the corrected energy and clear sky conditions. The measured flux is the ratio between this sum and the total observation time. 

We propose a similar algorithm. However, we do not use an event-wise shower maximum position ($H_{max}$) in our method because the reconstructed $H_{max}$ is biased by the presence of clouds (see Appendix B). We have used collection area obtained from the simulations  of cloudless conditions.
Moreover, the collection area is calculated as a function of reconstructed energy, i.e., it is defined as $A_{eff, rec}(E_{rec}, T, H) = A_0 dN_{surv}(E_{rec}, T, H) / dN_{sim}(E_{rec})$, where $A_0$ is the total simulated area, $dN_{surv}(E_{rec}, T, H)$ is the number of $\gamma$ rays surviving all the cuts with reconstructed energy between $E_{rec}$ and $E_{rec}+dE_{rec}$ and $dN_{sim}(E_{rec})$ is the number of simulated events with true energy between $E_{rec}$ and $E_{rec}+dE_{rec}$.
In the calculations of $A_{eff, rec}(E_{rec}, T, H)$, the energy spectrum of the MC $\gamma$ rays was assumed to be Crab-like. 
Since the same spectrum was used in the calculations of $A_{eff, rec}(E_{rec}, T, H)$, the energy bias and resolution does not bias the reconstructed spectrum (at the assumption, that the energy migration of corrected energies scales in the same way as the corrected energy itself). 
In a realistic case of an unknown spectrum of the source, a similar approach can be used with the assumed spectral parameters obtained from a minimization procedure (forward folding) or unfolding in true energy can be applied \citep{alber2007}.  

In order to determine a spectrum from the data taken in the presence of clouds  each $\gamma$-like event selected from the source is treated in the following way.  
At the first step, the reconstructed energy is corrected based on the energy bias that corresponds to both the altitude and total transmission of the cloud (approximation by formula~\ref{eq:bias2}). The second step of our algorithm is exactly the same as in \citet{fruck13,fruck15}. For each reconstructed event, we use the effective collection area $A_{eff,rec}(E_{corr}\cdot \tau_{A} (E_{corr},T,H),1,0)$.
Note that in our method, both variables ($E_{corr}$ and $\tau (E_{corr},T,H)$) are described by the same physical function: fraction of the photons created above the cloud altitude. Therefore, the method can be easily implemented for all possible heights of clouds. The only limit of our method is the total transparency of the additional extinction layers - for data taken with cloud the transmission should be higher than 0.5.

It is worth mentioning here, that we considered a simple case of a single-layer cloud that does not change over time, but the method can be extended to more complicated cases of cloud transmission variable in time. The total observation time should be split into parts in which the cloud can be considered as stable. Next, the correction should be applied to each part, separately computing the effective area for each bin of the energy and time. The final spectrum can be estimated using the average collection area weighted with the effective observation time in each time bin (see also \citealp{fruck15}).The time binning is limited by the LIDAR measurements accuracy (in order of minutes \citealt{fruck13}).

In case of a potential multiple cloud layers, the total atmospheric transmission can be estimated based on $F_{ab}$ using a modified formula which takes into account an altitude dependent extinction correction. One could divide the atmosphere in $M$ bins of the altitude ($h$) and calculate the total atmospheric transmission for the shower at energy $E$ ($\tau(E)$) as:

\begin{equation}
  \tau(E) = \sum_{i=1}^M (T_{below}((h_{i+1}+h_i)/2) \left( F_{ab}(E, h_i) - F_{ab}(E, h_{i+1}) \right)
  \label{eq:multi}
\end{equation}
where T$_{below}(h)$ is the total transmission of clouds below the altitude of $h$. We expect that the impact of the multiple cloud layers on the $\gamma$/hadron efficiency is smaller than in case of a single layer of cloudy medium. The feasibility of this approach will be the subject of future study.

\begin{figure}[!htb]
  \begin{center}
    \includegraphics[width=0.36\textwidth]{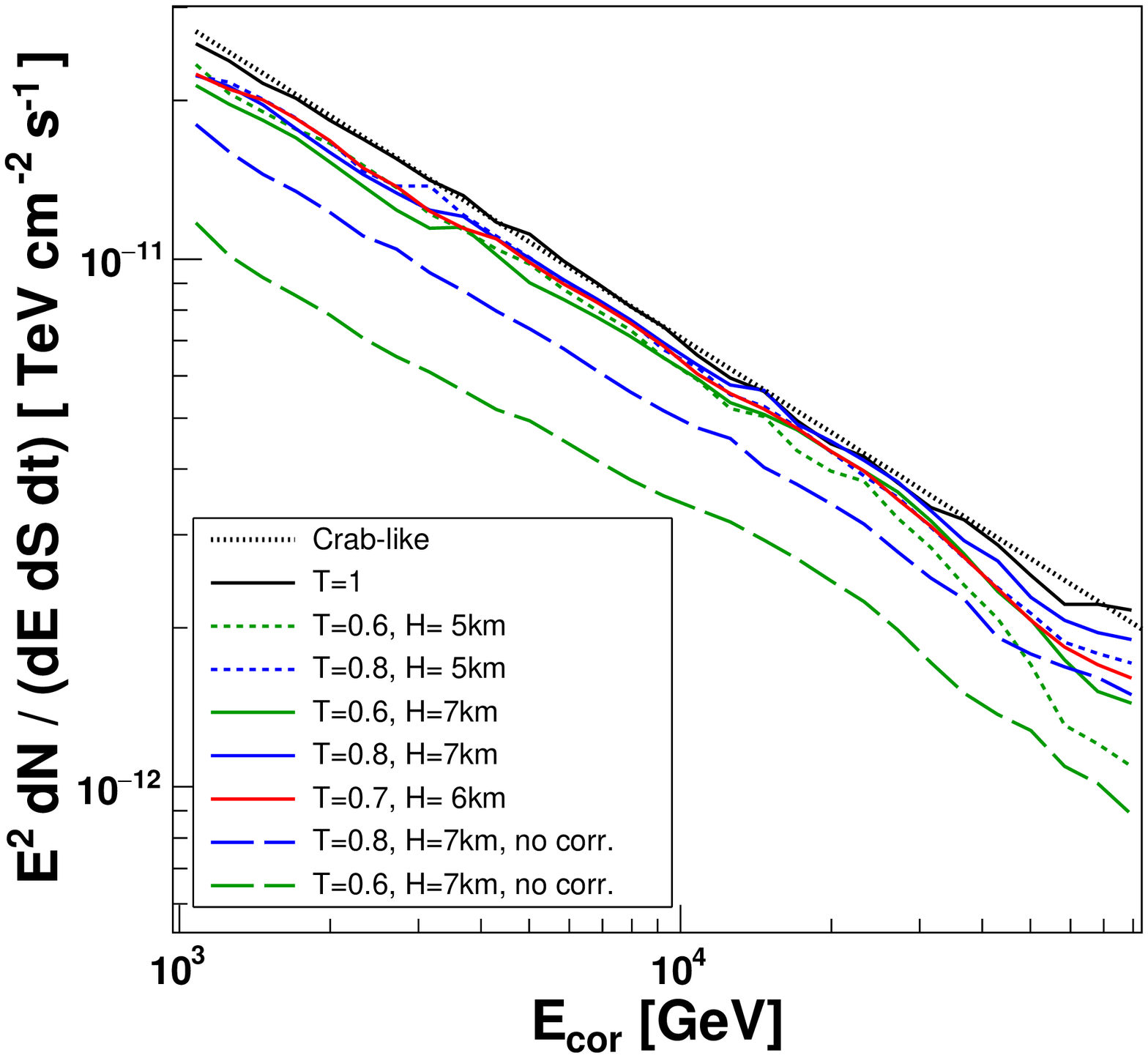}
    \includegraphics[width=0.36\textwidth]{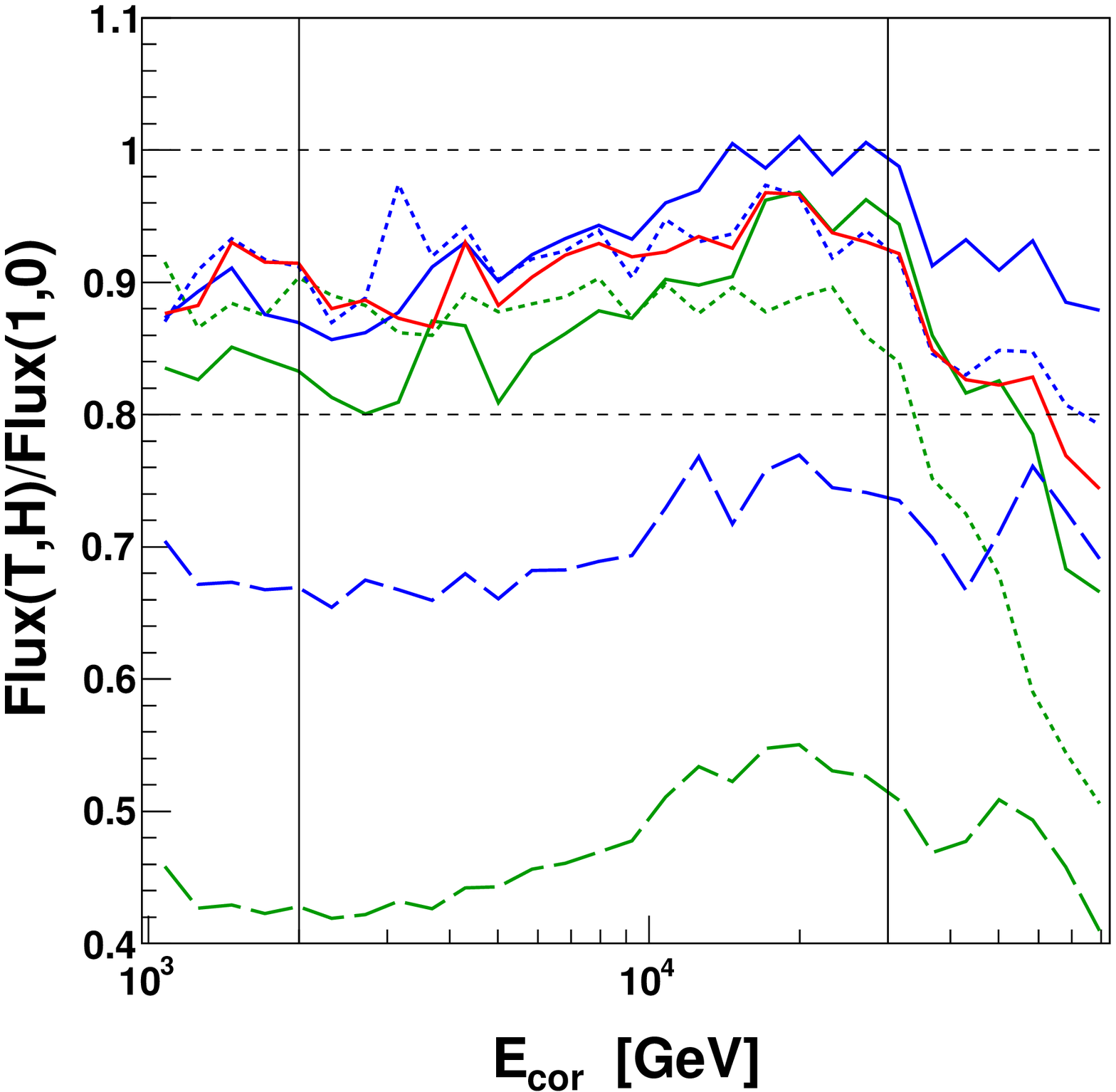}
    \includegraphics[width=0.36\textwidth]{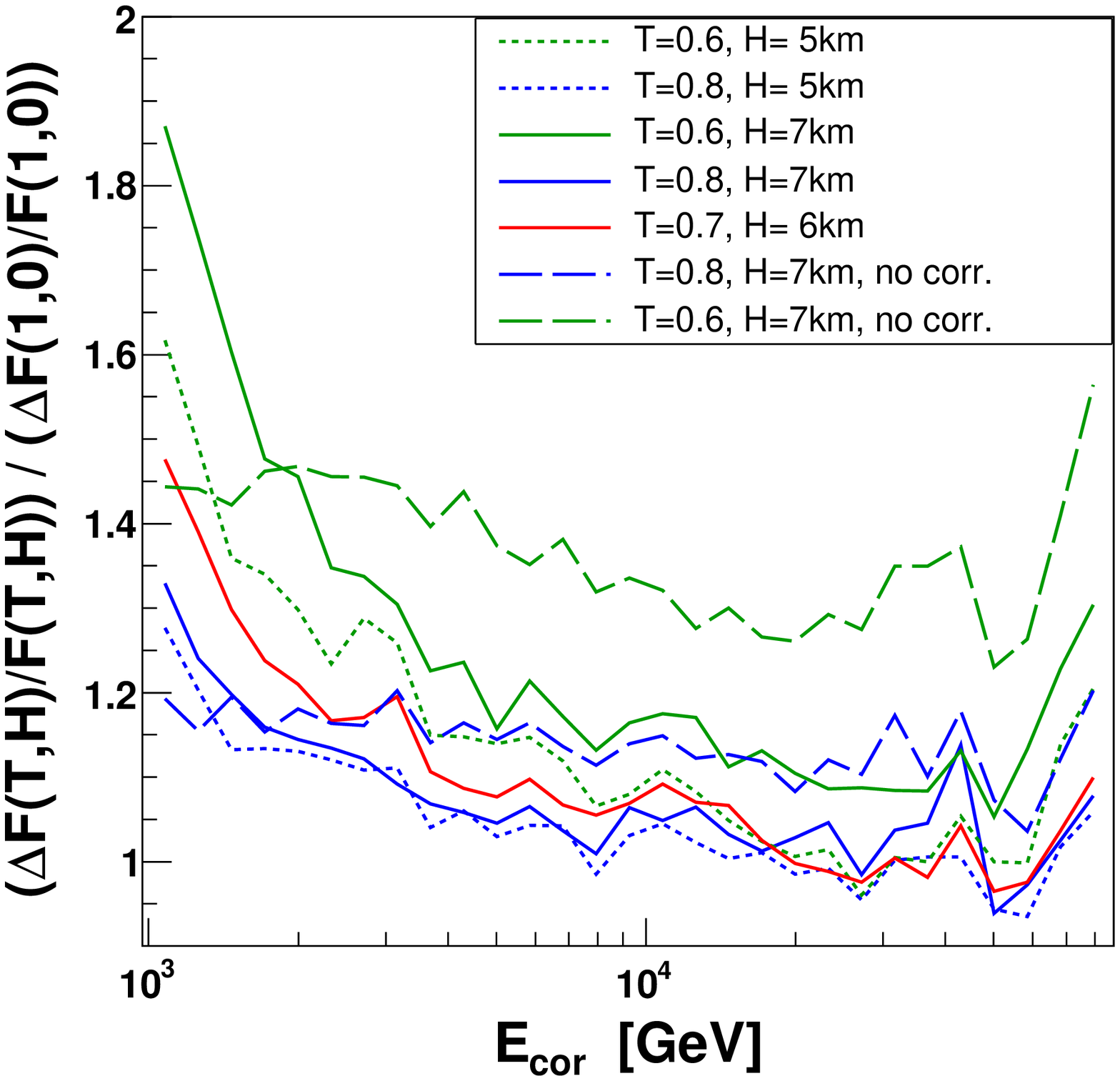}
    \caption{Top panel: A Crab-like spectra obtained for clear sky (black solid line) and under the presence of clouds (see legend). All short-dashed curves for T$<$1 show the results of the analysis which include the energy correction and the collection area approximation (described in text). Long-dashed lines show cloud simulation results obtained without any corrections (in this case $E_{cor}$ is $E_{rec}$).The black dotted line corresponds to a Crab-like spectrum that was used for the MC normalization. Middle panel: The ratio between the reconstructed flux 
and the flux obtained from full MC simulation for the cloudless condition. Bottom panel: The ratio of relative statistical uncertainties of the reconstructed flux for observation with clouds to the one expected for a clear sky.}
\label{fig:flux}
  \end{center}
\end{figure}

The source spectra reproduced using our method are presented as color solid and short-dashed lines in the top panel of Figure~\ref{fig:flux}. Only the black solid line (no cloud) presents the results obtained by using the effective collection area from full Monte Carlo simulations. In cases of T$<$1 we have applied $A_{eff,rec}(E_{corr}\cdot\tau_{A} (E_{corr},T,H),1,0)$ after $\gamma/hadron$ separation. The black dotted line corresponds to a Crab-like spectrum that was used for the MC normalization. Long-dashed lines show cloud simulation results obtained without any corrections.  
It is seen in the figure that between 1~TeV and 80~TeV using the full Monte Carlo chain in data analysis leads to proper flux reconstruction for the clear sky condition. In the previous subsections we discussed the variables: angular resolution, energy resolution, effective collection area after reconstruction and hadroness cut efficiency. All of them influence the reconstructed spectrum. However, the distribution of the first one is crucial in a choosing of $\theta^{2}$ cut. This cut influences both the number of events selected as the $\gamma$s from the source and collection area after $\gamma/hadron$ separation.
Additionally, the angular resolution is well-described by our approach for cloud transparencies above 0.4. Furthermore, for E$\ge$~2~TeV and T$\geq$0.6 the resolution  is almost independent of energy. Thus the $\theta^{2}$ cut should not change the reconstructed spectra.
Note that the presence of clouds only slightly affects the energy resolution for energies above 2~TeV. It is relatively good even in the case of clouds at 7~km with T=0.6 where the resolution is $\sim12\%$. Therefore, the impact of energy dispersion on the spectral determination should be similar in the case of cloudless sky and observations in the presence of clouds.
  
The middle panel of Figure \ref{fig:flux} presents the ratio between the reconstructed flux by using the correction method and the flux obtained from full MC simulation of the cloudless condition. This ratio demonstrates the accuracy of our method.
For energies below $\sim$2~TeV, where the threshold effects are dominant, the flux corrected for the presence of clouds is underestimated by less than 20$\%$.

From $\sim$2~TeV up to $\sim$30~TeV, the spectra reconstructed from data with a cloud at 5 km are almost proportional to the results for the clear sky, i.e. in this case only the flux normalization is affected, not the reconstruction of the spectral index. Furthermore, the almost constant flux underestimation is greater for lower transparencies (compare of the green and blue lines in this figure). In addition, the altitude of the cloud also influences the underestimation. 
It is worth mentioning that only in case of a cloud with T=0.6 at 5~km a.s.l. (see green dashed line) the efficiency of the  $Hadroness$ selection does decrease with energy, which results in a slightly faster degradation of the presented ratio with energy.

For energies above 30~TeV, the flux underestimation increases with energy mainly due to the fast degradation of the hadroness cut efficiency (see Figure ~\ref{fig:fractionG95ii}). Moreover the spectral index cannot be properly determined in this energy range due to the fact that the presented ratio changes with energy. Only very transparent (T=0.8) and high (7~km) clouds can be analyzed with the 20$\%$ accuracy of the method. Thus the method should not be used for energies above $\sim$30~TeV in case of lower cloud transparencies or lower cloud altitudes.

The bottom panel of Figure \ref{fig:flux} shows the ratio of relative statistical uncertainties of the reconstructed flux for Crab-like source in the case of observations with clouds to the one expected for a clear sky.
In most of the investigated cases the relative increase of the relative statistical uncertainties is small, $\lesssim20\%$. 

\section{Conclusions}
We have studied a correction method for data taken by SST-1M telescopes in the presence of clouds for a hypothetical Armazones site. For this purpose we have used the standard CTA simulation software $sim\_telarray$ and MARS/Chimp analysis chain. To simulate additional extinction by a cloud layers, the standard atmospheric extinction file has been modified for each studied cloud altitude and transmission.

The correction method we propose does not require a dedicated MC simulation for analysis of the data taken in the presence of clouds. To use the correction, one needs to obtain, e.g. typically based on LIDAR data, the height and the total transmission of the cloud. Next, find (based on CORSIKA simulation) the dependence of the fraction of the Cherenkov light created above the cloud to all produced photons (both counted for impact parameter of the photons $>$80~m) on the primary energy of the $\gamma$-ray for the chosen observation site.

The validity range of the method is limited by two effects. First, threshold effects - we estimate that below $\sim$ 2-3~TeV the degradation of the trigger and reconstruction rates due to the presence of clouds are not described well enough by the scaling energy. Furthermore, both the angular and energy resolutions are much worse close to the threshold. Second, the deformation of the shower images at very high energies (above $\sim$ 30~TeV) for clouds at 7~km a.s.l. or higher significantly affects the hadroness cut efficiency while using values that are optimized for clear sky (we do that to avoid time consuming full MC simulations for different cloud altitudes and transparencies).

For the presence of clouds at an altitude H, the energy bias can be described by the fraction of photons created above the cloud and the bias curve obtained from cloudless conditions. The energy bias approximation we propose works for cloud transmissions above $\sim 0.5$. Similarly, the angular resolution can be predicted by scaling the energy by the total atmospheric transmission for the cloud transparencies $\geq$0.4. Using the same scaling for the effective collection area after reconstruction results in an overestimation of the calculated collection area (by less than 20$\%$) in comparison to that obtained from full MC for energies above $\sim$~2~TeV if the cloud transmissions are higher than 0.4 and $\sim 0.6$ for altitude 5 and 7~km respectively.

The systematic uncertainty of the correction method presented for the reconstructed spectra is smaller than $\sim$~20$\%$ in the energy range between 2 and 30~TeV. For higher energies the uncertainty of the method is dominated by the $Hadroness$ cut efficiency. We estimate that one may safely use a correction for cloud transmissions $\geq$ 0.6. The presented method was tested for cloud altitudes between 5 and 7 km a.s.l..
For higher clouds the proposed correction is smaller as it is based on the fraction of Cherenkov light created above a cloud (see Fig.~\ref{fig:fraction} in Appendix A). This is in agreement with the results presented by \citep{sob14}: for high energies smaller impact of the clouds located higher is expected. For lower clouds the correction simplifies as nearly all Cherenkov photons are attenuated by a cloud in the same way, resulting in less deformed images.

We conclude that SST-1M telescopes can be efficiently used for observation in the wide range of clouds. Such data can be analyzed using the simple method presented in this paper if both the energy and cloud transmission are within the limits mentioned above.

\section*{Acknowledgements}
This work is supported by the National Science Centre grant No. UMO-2016/22/M/ST9/00583. We would like to thank CTA Consortium and MAGIC Collaboration for allowing us to use their software, SST-1M sub-consortium for providing us the prod3 parameters of the SST-1M telescopes and M. Gaug and A. Moralejo for helpful discussions and anonymous journal referees for their valuable comments to the manuscript.

This paper has gone through internal review by the CTA Consortium.

\section*{Appendix A. Fraction of Cherenkov photons created above the cloud}

\begin{figure}[!htb]
\begin{center}
  \includegraphics[width=0.45\textwidth]{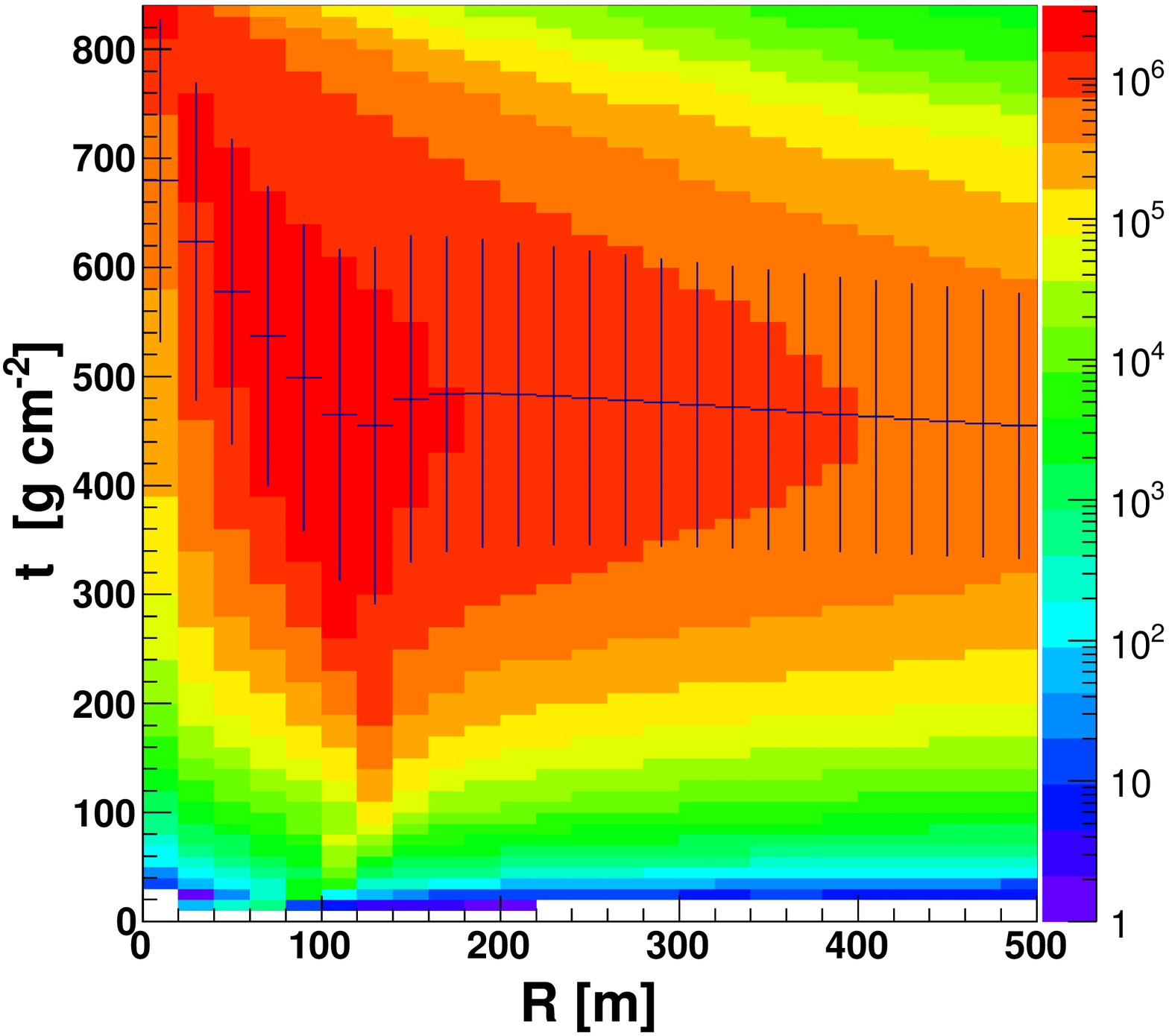}\\
  \includegraphics[width=0.45\textwidth]{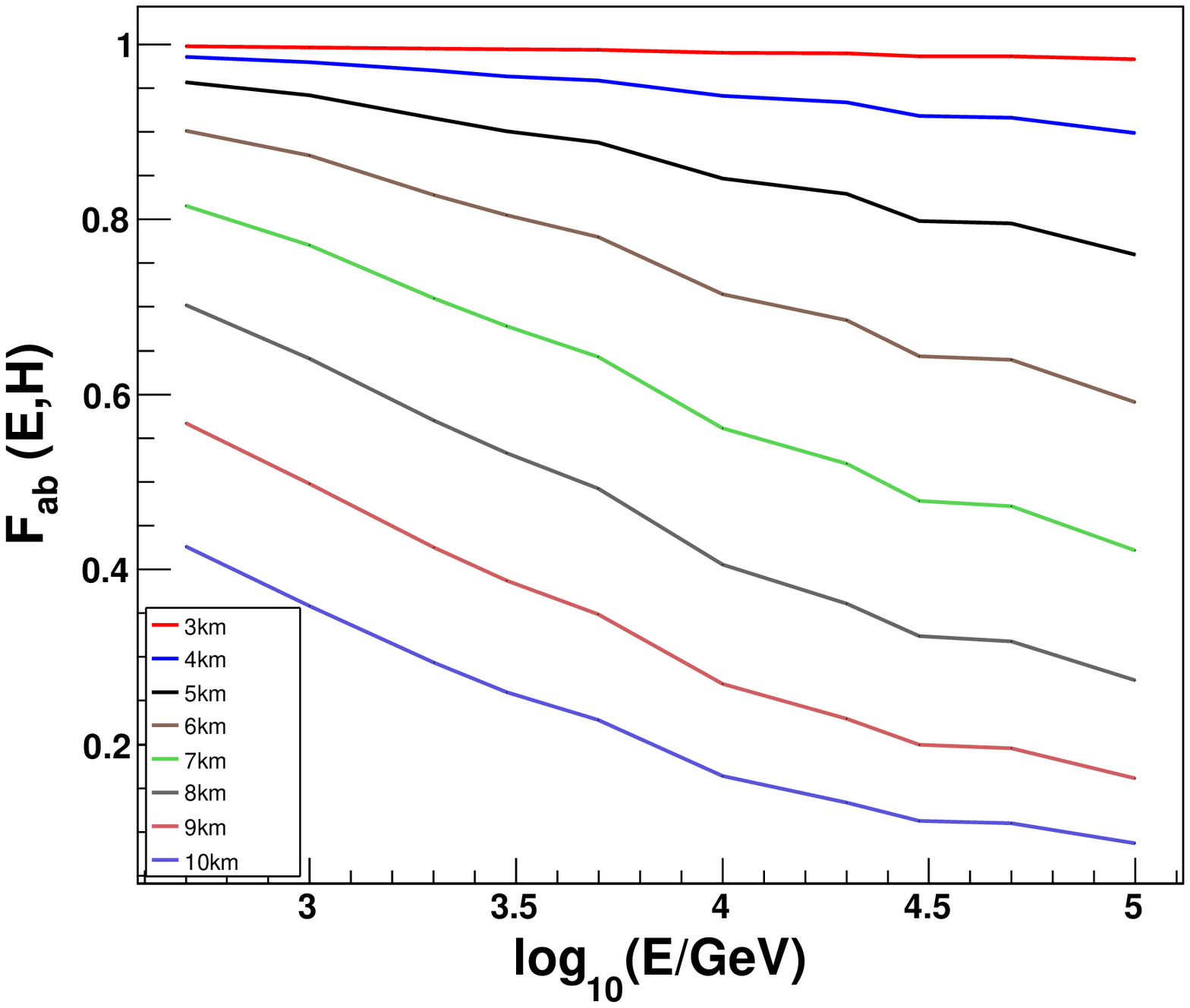}  
\caption{Top panel: The average number of produced photons versus both depth of its production (in $g/cm^{2}$) and the impact parameter for the primary energy of 20~TeV. Bottom panel: Mean fraction of Cherenkov photons produced above different cloud level versus $log_{10}(E)$. $F_{ab} (E,H)$ was calculated for photons that hit the ground at distances higher than 80~m for observation level 2500~m a.s.l.. The standard deviation (shower to shower fluctuations) of the fraction distribution is below 0.12 for each simulated energy and altitude}
\label{fig:fraction}
\end{center}
\end{figure}

The fraction of all Cherenkov photons produced above the cloud has been calculated based on additional Monte Carlo simulations using CORSIKA. The primary $\gamma$-rays with fixed energies were simulated in order to check the distributions of the Cherenkov light production altitude (more precisely corresponding thickness of the atmosphere) for a given distance from the core axis (R). The top panel of Figure \ref{fig:fraction} shows the average number of photons versus both  its production depth (in $g/cm^{2}$) and the distance from the core for the primary energy of 20~TeV. The parameters describing the light production depth distribution (presented as profiles in this figure) are very similar for R equal or higher than 80~m in all simulated energies. The stereo trigger of our IACTs array requires that at least one triggered telescope is located at a distance higher than 100~m from the shower axis core. Thus we conclude that impact distances R$>$80~m play dominant role in the energy reconstruction. Before integrating the presented distribution to the total amount of Cherenkov light that hits the ground above this distance we applied a simple correction for Rayleigh scattering that depends on both the photon production and observation altitudes. The ratios between the total number of photons created above different cloud altitudes and the total amount of the produced Cherenkov light (both for R$>$80~m) versus the energy of $\gamma$-ray are presented in the bottom panel of Fig.~\ref{fig:fraction}. as a function of the primary energy. Those fractions ($F_{ab} (E,H)$) can be fitted as a function of the energy and they are used in the bias approximation~\ref{eq:bias2}.

Based on the $F_{ab} (E,H)$ one can estimate the total atmospheric transmission for the $\gamma$-ray with energy E for a cloud at the altitude H with the total cloud transparency equal to T (see formula \ref{eq:tau})

However, for the energy scaling we use the corrected total atmospheric transmission that includes a constant A:
\begin{equation}
  \tau_{A}(E,T,H) \equiv 1 - A \cdot(1-T)\cdot F_{ab}(E,H)
\label{eq:tauwitha}
\end{equation}

The factor A is applied in the same way as in the bias approximation and it is equal to the same value of 1.2 in all  results presented in this paper.
By using $\tau_{A}$ instead of simply $\tau$ in the formulas \ref{eq:bias2}, \ref{eq:angular} and~\ref{eq:effcollarea} we get better agreement between the full MC results and proposed approximations in Figures \ref{fig:bias}, \ref{fig:angular} and \ref{fig:collarea}. 

\section*{Appendix B. The Effect of the presence of clouds on the shower maximum reconstruction}

\begin{figure}[!htb]
  \includegraphics[width=0.45\textwidth]{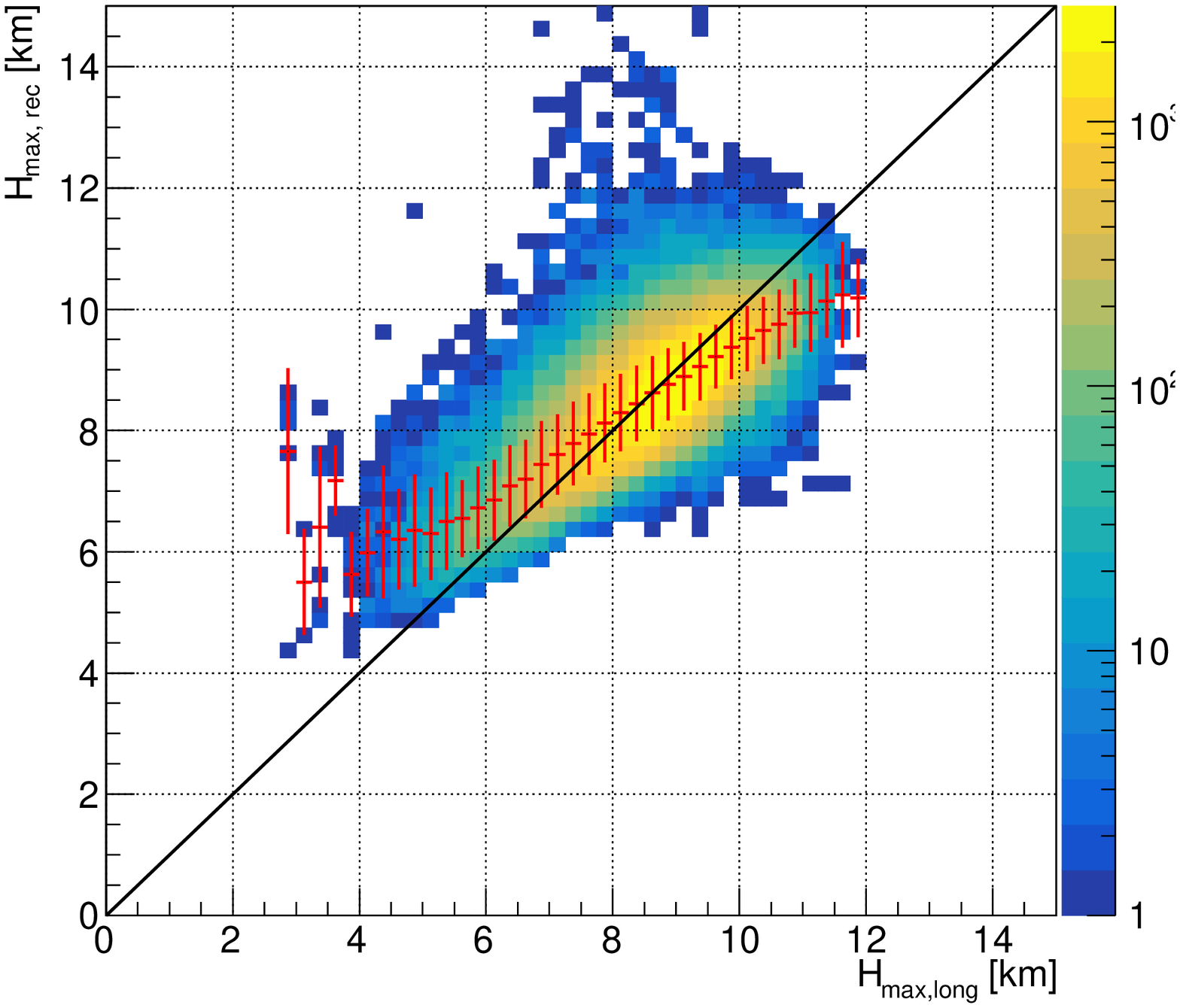}
  \includegraphics[width=0.45\textwidth]{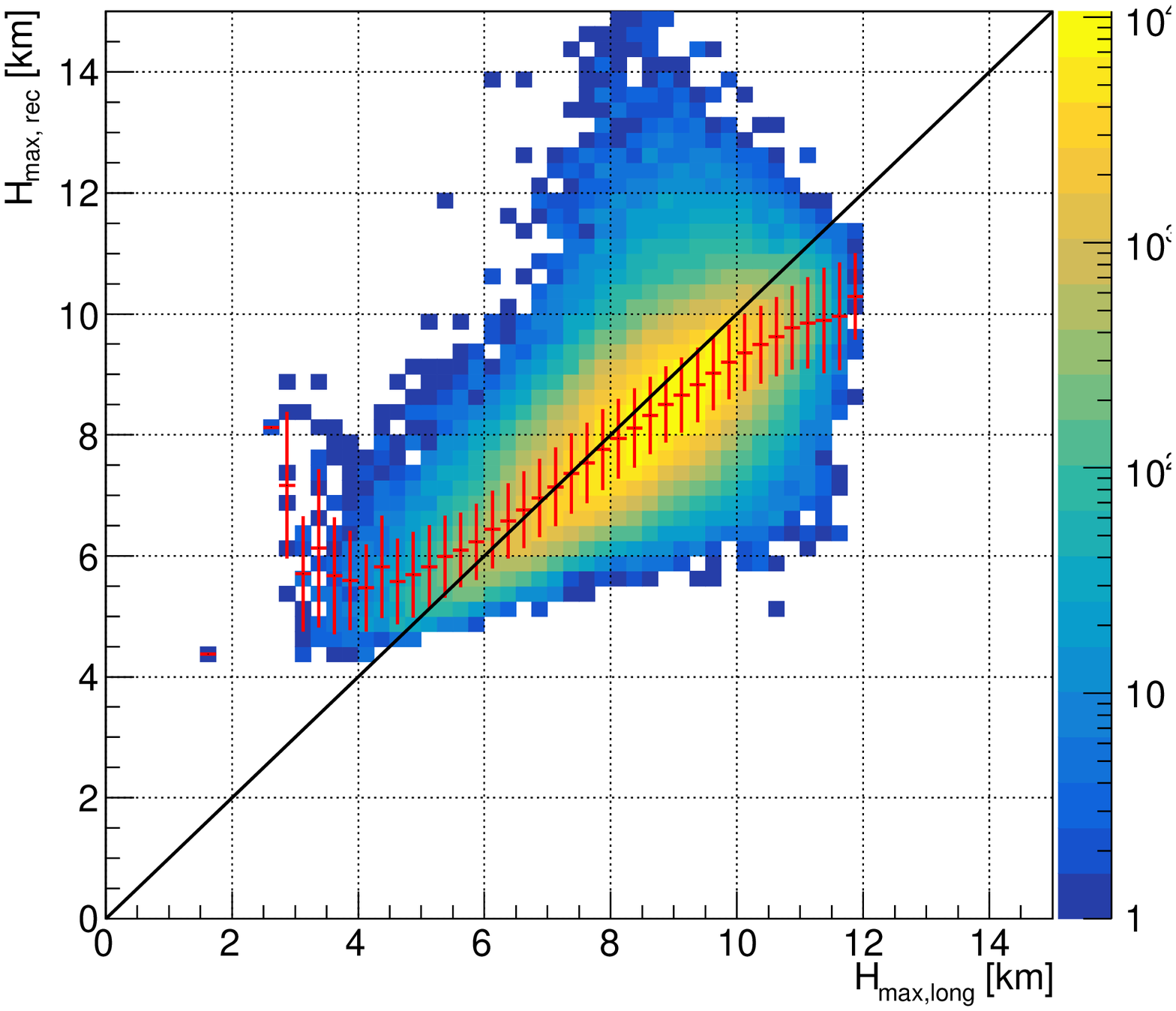}
  \includegraphics[width=0.45\textwidth]{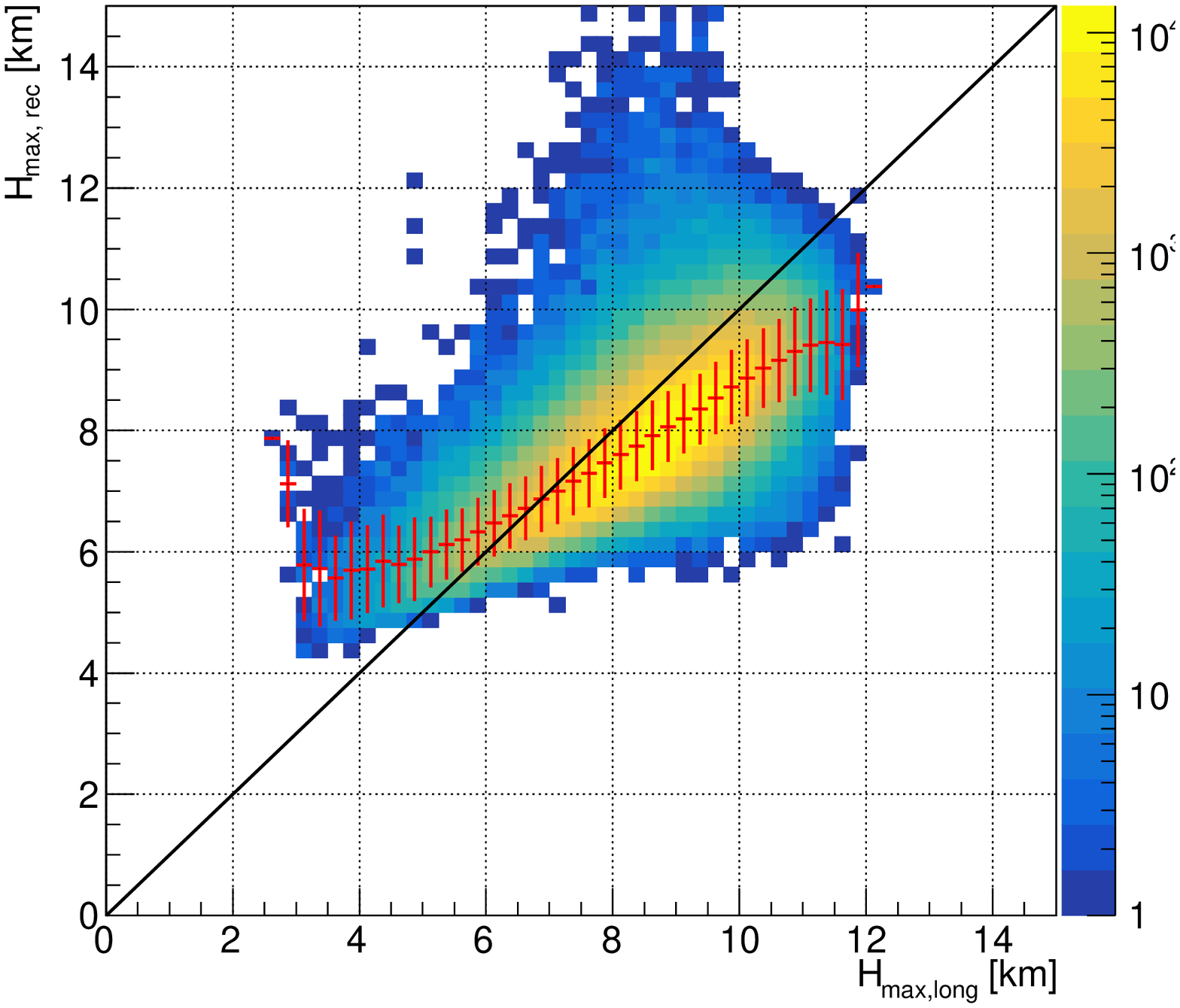}
  \caption{Reconstructed height of the shower maximum as the function of the height of the shower maximum obtained from the longitudinal distribution of particles. Only events with energy between 3 and 30 TeV are used. G95 and $\theta^2<$0.025~deg$^{2}$ cuts are applied. The top panel is for cloudless conditions, the middle panel is for cloud with transmission of 0.6 at 5\,km a.s.l., and the bottom panel for such a cloud at the height of 7\,km a.s.l.}\label{fig:hmax}
\end{figure}

In Fig.~\ref{fig:hmax} we investigate the effect of the cloud on the reconstruction of the shower maximum.
In the case of cloudless conditions the height of the shower maximum reconstruction has nearly no bias for showers with shower maximum at about 9\,km a.s.l..
For showers that fluctuated very high or very low in the atmosphere there is an opposite bias in the reconstruction.
Events which develop at high altitude will have significant absorption of the UV light which will bias the reconstructed shower maximum to lower values.
In contrary the events that develop deep in the atmosphere will have their tails cut due to limited FoV of Cherenkov telescopes and angular distribution of Cherenkov light, biasing the height of the shower maximum to higher values.
In the case of a low cloud at 5\,km a.s.l. the whole distribution is slightly biased to the lower values and in addition a higher spread is seen.
In the case of a higher cloud at 7\,km a.s.l. much stronger bias is visible if the height of the shower maximum is above the cloud.
In such a case the cloud can cut the shower in half affecting its stereoscopic reconstruction.
The events with the true height of the shower maximum of about 8.5\,km a.s.l. but reconstructed at the height of about 12\,km a.s.l. are several tens of TeV showers observed at large impact parameter that are highly misreconstructed most probably due to angular distribution of the observed light.


\section*{References}

\begin{thebibliography}{00}

\bibitem[Actis et al.(2011)]{act11} Actis M et al.\ 2011, Experimental Astronomy 32, 193 
\bibitem[Acharaya et al.(2013)]{ach13} Acharaya B. et al.\ 2013, Astropart. Phys. 43, 3
\bibitem[Acharyya et al.(2019)]{ach19} Acharyya A. et al.\ 2019, Astropart. Phys. 111, 35
\bibitem[Adamczyk $\&$ Sobczy\'nska(2017)]{adam16} Adamczyk K. $\&$ Sobczy\'nska D.\ 2017 AIP Conference Proceedings 1792, 080013
\bibitem[Aharonian et al.(2004)]{aha06} Aharonian F. et al.\ 2004, A\&A 457, 899
\bibitem[Albert et al.(2008)]{alber2008} Albert et al.\ 2008, Nuclear Instruments and Methods in Physics Research A 588, 424
\bibitem[Albert et al.(2007)]{alber2007} Albert et al.\ 2007, Nuclear Instruments and Methods in Physics Research A 583, 494
\bibitem[Aleksi\'{c} et al.(2016)]{ale16} Aleksi\'{c} J. et al.\ 2016, Astropart. Phys. 72, 76
\bibitem[Aleksi\'{c} et al.(2012)]{ale12} Aleksi\'{c} J. et al.\ 2012, Astropart. Phys. 35, 435
\bibitem[Barnacka et al.(2013)]{barnacka13} Barnacka A et al. \ 2013 Proc. 33th Int. Cosmic Ray Conf. (Rio de Janerio), 2013arXiv1307.3409B
\bibitem[Bass et al.(1998)]{urq1} Bass S. A. et al.\ 1998, Prog.Part.Nucl.Phys. 41, 225
\bibitem[Bernl\"{o}hr(2000)]{bern00} Bernl\"{o}hr K.\ 2000, Astropart. Phys. 12, 255 
\bibitem[Bernl\"{o}hr(2008)]{bern08} Bernl\"{o}hr K.\ 2008, Astropart. Phys. 30, 149 
\bibitem[Bernl\"{o}hr(2013)]{bern13} Bernl\"{o}hr K.\ 2014, Proc. of the 1st AtmoHEAD workshop (2013), arXiv:1402.5081
\bibitem[Bleicher et al.(1999)]{urq2} Bleicher M. et al.\ 1999, J.Phys.G: Nucl. Part. Phys. 25, 1859
\bibitem[Chitnis \& Bhat(1998)]{chitnis98} Chitnis V.R. \& Bhat P.N.\ 1998, Astropart. Phys. 9, 45 
\bibitem[ Devin et al.(2019)]{devin} Devin J. et al.\ 2019, EPJ Web of Conferences Vol. 197, 01001
\bibitem[Dorner et al.(2009)]{dorn09} Dorner D. et al.\ 2009, A\&A 493, 721
\bibitem[Doro et al.(2013)]{doro13} Doro M. et al.\ 2014, Proc. of the 1st AtmoHEAD workshop (2013), arXiv:1402.0638
\bibitem[Fruck et al.(2013)]{fruck13} Fruck C. et al.\ 2013, Proc. 33th Int. Cosmic Ray Conf. (Rio de Janeiro 2013), arXiv:1403.3591
\bibitem[Fruck \& Gaug(2015)]{fruck15} Fruck C. \& Gaug M.\ 2015, Proc. of AtmoHEAD 2014, EPJ Web of Conferences, Vol. 89, id.02003
\bibitem[Garrido et al.(2013)]{garrido13} Garrido D. et al.\ 2013 Proc. 33th Int. Cosmic Ray Conf. (Rio de Janeiro 2013), 0465
\bibitem[Gaug(2016)]{gaug16} Gaug M. \ 2017, Proc. of AtmoHEAD workshop (2016), EPJ Web of Conferences Vol. 144, 01003
\bibitem[Gaug et al.(2018)]{gaug18} Gaug M. et al.\ 2019, Proc. of AtmoHEAD workshop (2018), EPJ Web of Conferences Vol. 197, 02005 
\bibitem[Hahn et al.(2014)]{hahn14} Hahn J. et al.\ 2014, Astropart. Phys. 54, 25
\bibitem[Heck et al.(1998)]{heck} Heck D. et al.\ 1998 Technical Report FZKA 6019 (Forschungszentrum Karlsruhe) 
\bibitem[Heck \& Pierog(2011)]{knapp} Heck D. \& Pierog T.\ 2011, EAS Simulation with CORSIKA: A Users Manual
\bibitem[Heller et al.(2017)]{heller17} Heller M. et al.\ 2017, Eur. Phys. J. C 77, 47
\bibitem[Holder et al.(2011)]{hold11} Holder J. et al.\ 2011, Proc. 32th Int. Cosmic Ray Conf. (Beijing) 12, 137
\bibitem[Hildebrand et al.(2013)]{hildebrand13} Hildebrand D. et al.\ 2013, Proc. 33th Int. Cosmic Ray Conf. (Rio de Janerio) p.3020
\bibitem[Hildebrand et al.(2017)]{hildebrand17} Hildebrand D. et al.\ 2017, Proc. 35th Int. Cosmic Ray Conf.(Busan, Korea 2017), id.779
 
\bibitem[Iarlori et al.(2016)]{iarl16}Iarlori M. et al.\ 2017, Proc. of AtmoHEAD workshop (2016), EPJ Web of Conferences Vol. 144, 01008
\bibitem[Kokhanovsky(2004)]{kokh08} Kokhanovsky A.\ 2004 Earth-Science Review 64, 189-241
\bibitem[Kurlandczyk \& Sarazin (2007)]{Kurlandczyk2007} Kurlandczyk H. \& Sarazin M.\ 2007 Proceedings of the SPIE ``Remote Sensing of Clouds and the Atmosphere XII'' Vol. 6745, id. 674507
\bibitem[L\'opez et al.(2013)]{lop13} L\'opez Oramas A.\ 2013, Proc. 33th Int. Cosmic Ray Conf. (Rio de Janerio), p0210, arXiv:1307.5092L
\bibitem[Nolan et al.(2010)]{nolan10} Nolan S. J. et al.\ 2010, Astropart. Phys. 34, 304
\bibitem[Ostapchenko(2006a)]{ostap06a} Ostapchenko S.\ 2006, Phys.Lett. B 636, 40
\bibitem[Ostapchenko(2006b)]{ostap06b} Ostapchenko S.\ 2006, Phys.Rev. D 74, 014026
\bibitem[Ostapchenko(2007)]{ostap07} Ostapchenko S.\ 2007, Collicers to Cosmic Rays, AIP Conference Proceedings 928, 118
\bibitem[Rulten et al.(2013)]{rult13} Rulten C. B. et al.\ 2014 Proc. of the 1st AtmoHEAD workshop (2013), arXiv:1403.2218
\bibitem[Sitarek et al.(2018)]{sit18} Sitarek J. et al.\ 2018, Astropart. Phys. 97, 1
\bibitem [Sliusar et al.(2017)]{sliusar17} Sliusar V. et al.\ 2017 Proc. 35th Int. Cosmic Ray Conf. (Busan, Korea 2017), arXiv:1709.04244
\bibitem[Sobczy\'nska(2009)]{sob09} Sobczy\'nska D. \ 2009, J.Phys.G: Nucl. Part. Phys. 36, 045201
\bibitem[Sobczy\'nska \& Bednarek(2013)]{sob13} Sobczy\'nska D. \& Bednarek W.\ 2013, Proc. 33th Int. Cosmic Ray Conf. (Rio de Janerio), 00335
\bibitem[Sobczy\'nska \& Bednarek(2014)]{sob14} Sobczy\'nska D. \& Bednarek W.\ 2014, J.Phys.G: Nucl. Part. Phys. 41, 125201
\bibitem[Sobczy\'nska \& Bednarek(2015)]{sob15} Sobczy\'nska D. \& Bednarek W.\ 2015, Proc. of AtmoHEAD (2014), EPJ Web of Conferences Vol. 89, id.03009
\bibitem [Weekes et al.(1989)]{whipple} Weekes T. C. et al.\ 1989, Astrophys. J. 342, 379
\bibitem[Weekes et al.(2002)]{weekes2002} Weekes T. C. et al.\ 2002 Astropart. Phys. 17, 221 
\bibitem[Valore et al.(2017)]{valore17} Valore L. et al.\ 2017 Proc. 35th Int. Cosmic Ray Conf. (Busan, Korea 2017), id 833
\bibitem[Zanin et al.(2013)]{zan13} Zanin R. et al.\ 2013, Proc. 33th Int. Cosmic Ray Conf. (Rio de Janerio), id. 773

\end{thebibliography}
\end{document}